\documentclass[11pt]{article}
\usepackage[left=1in, right=1in, top=1in]{geometry}
\usepackage{hyperref}
\usepackage{xargs}
\usepackage{natbib}
\usepackage{fancyhdr}
\usepackage{setspace}
\usepackage{lastpage}
\usepackage{upgreek}
\usepackage{graphicx}	% Including figure files
\usepackage[lofdepth,lotdepth]{subfig}
\usepackage{amsmath,mathtools,amsthm}	% Advanced maths commands
\usepackage{amssymb,dsfont,bbm}	% Extra maths symbols
\usepackage[ruled,vlined]{algorithm}  % algorithm
\usepackage{xcolor}  % colors
\usepackage{comment}
\usepackage[textwidth=3cm,textsize=footnotesize]{todonotes}
\usepackage{bm}

\setcitestyle{authoryear}

\usepackage{aliascnt}
\usepackage{cleveref}
\newcommand{\frobnorm}[1]{\left\Vert #1 \right\Vert_{\mathrm{F}}}
\newcommand{\normop}[1]{\left\lVert#1\right\rVert}
\newcommand{\ps}[2]{\langle#1,#2 \rangle}

\def\ie{i.e.}
\def\param{\boldsymbol{\beta}}
\def\Param{\mathcal{B}}
\def\hparam{\hat{\param}}
\newcommand{\estQ}[2]{Q_{#1,#2}}
\newcommand{\testQ}[2]{\tilde{Q}_{#1,#2}}
\newcommand{\estalpha}[2]{\bar{\alpha}_{#1,#2}}
\def\mcg{\mathcal{G}}
\def\Id{\mathrm{I}}
\def\rmd{\mathrm{d}}
\def\rme{\mathrm{e}}
\def\rset{\mathbb{R}}
\def\NtrainPath{T}

\def\Jac{\operatorname{J}}
\def\ltwo{\mathrm{L}^2}

\newtheorem{exmp}{Example}[section]

\newtheorem*{excont}{\bf{Example \continuation}}
\newcommand{\continuation}{??}
\newenvironment{continueexample}[1]
 {\renewcommand{\continuation}{\ref{#1}}\excont[\bf{continued}]}
 {\endexcont}

\newcommand{\jac}[1]{\operatorname{J} _{#1}}
\def\eqsp{\,}

\newcommand{\diag}{\operatorname{diag}}

\newcommand{\indiacc}[1]{\mathbbm{1}_{\{#1\}}}
\def\P{\mathsf{P}}
\def\C{\bar{C}}
\def\PE{\mathsf{E}}
\def\PVar{\mathsf{Var}}
\def\PCov{\mathsf{Cov}}
\newcommandx{\CPE}[3][1=]{\mathsf{E}_{#1}\left[\left. #2 \, \right| #3 \right]}
\def\ntrunc{n_0}
\def\qtrunc{b_0}

\def\W{\mathsf{W}}
\def\X{\mathsf{X}}
\def\nset{\mathbb{N}}
\def\rset{\mathbb{R}}
\def\kind{\mathbf{k}}
\def\lind{\bm{\ell}}

\def\eind{\mathbf{e}}
\def\rind{\mathbf{r}}
\def\qind{\mathbf{q}}

\def\bigK{\mathbf{K}}
\def\basis{\psi}

\newcommand{\chunk}[3]{#1_{#2:#3}}
\newcommand*{\const}{\mathrm{const}}

\DeclareMathAlphabet\mathbfcal{OMS}{cmsy}{b}{n} % Bold Mathcal font
\DeclareMathOperator*{\argmin}{argmin}

\newtheorem{theorem}{Theorem}
\crefname{theorem}{theorem}{Theorems}
\Crefname{Theorem}{Theorem}{Theorems}

\newaliascnt{lemma}{theorem}
\newtheorem{lemma}[lemma]{Lemma}
\aliascntresetthe{lemma}
\crefname{lemma}{lemma}{lemmas}
\Crefname{Lemma}{Lemma}{Lemmas}

\newaliascnt{corollary}{theorem}
\newtheorem{corollary}[corollary]{Corollary}
\aliascntresetthe{corollary}
\crefname{corollary}{corollary}{corollaries}
\Crefname{Corollary}{Corollary}{Corollaries}

\newaliascnt{proposition}{theorem}
\newtheorem{proposition}[proposition]{Proposition}
\aliascntresetthe{proposition}
\crefname{proposition}{proposition}{propositions}
\Crefname{Proposition}{Proposition}{Propositions}

\newaliascnt{definition}{theorem}

\aliascntresetthe{definition}
\crefname{definition}{definition}{definitions}
\Crefname{Definition}{Definition}{Definitions}

\newaliascnt{definitionProposition}{theorem}

\aliascntresetthe{definitionProposition}
\crefname{Proposition and Definition}{Proposition and Definition}{Proposition and Definition}
\Crefname{Proposition and Definition}{Proposition and Definition}{Proposition and Definition}

\newaliascnt{remark}{theorem}
\newtheorem{remark}[remark]{Remark}
\aliascntresetthe{remark}
\crefname{remark}{remark}{remarks}
\Crefname{Remark}{Remark}{Remarks}

\crefname{example}{example}{examples}
\Crefname{Example}{Example}{Examples}

\crefname{figure}{figure}{figures}
\Crefname{Figure}{Figure}{Figures}

\newtheorem{assumption}{\textbf{H}\hspace{-3pt}}
\Crefname{assumption}{\textbf{H}\hspace{-3pt}}{\textbf{H}\hspace{-3pt}}
\crefname{assumption}{\textbf{H}}{\textbf{H}}

\Crefname{assumptionL}{\textbf{L}\hspace{-3pt}}{\textbf{L}\hspace{-3pt}}
\crefname{assumptionL}{\textbf{L}}{\textbf{L}}

\Crefname{assumptionG}{\textbf{G}\hspace{-3pt}}{\textbf{G}\hspace{-3pt}}
\crefname{assumptionG}{\textbf{G}}{\textbf{G}}

\begin{document}

% Title again
\title{Variance reduction for additive functional of Markov chains via martingale representations}

\author{D. Belomestny~\footnote{Duisburg-Essen University, Germany, and HSE University, Russia, \texttt{denis.belomestny@uni-due.de}. }, \, E. Moulines~\footnote{Ecole Polytechnique, France, and HSE University, Russia, \texttt{eric.moulines@polytechnique.edu}.}, and  S. Samsonov~\footnote{HSE University, Russia,  \texttt{svsamsonov@hse.ru}.}}
\maketitle

% Abstract
%\begin{abstract}
%hhh
%\end{abstract}

% example code and solutions can be found {\color{blue} \hyperlink{https://github.com/joshspeagle/XXX}{online}}

\begin{abstract}
In this paper we propose an efficient variance reduction approach for  additive functionals of Markov chains relying on a novel discrete time martingale representation. Our approach is fully non-asymptotic and does not require the knowledge of the stationary distribution (and even any type of ergodicity) or specific structure of the underlying density.  By rigorously analyzing the  convergence properties of the proposed algorithm, we show that its cost-to-variance product is indeed  smaller than one of the naive algorithm. The numerical performance of the new method is illustrated for the Langevin-type Markov Chain Monte Carlo (MCMC) methods.
\end{abstract}

%% Main body

\section{Introduction} \label{sec:intro}
Markov chains and Markov Chain Monte Carlo (MCMC) algorithms play a crucial role in modern numerical analysis, finding various applications in such research areas as Bayesian inference, reinforcement learning and online learning. As an illustration, suppose that we aim at computing $\pi(f):= \int f(x) \pi(\rmd x)$, where $f: \rset^d \to \rset$ is a function in $\ltwo(\pi)$ and $\pi$ has a smooth and everywhere positive density  w.r.t the Lebesgue measure (By abuse of notation, we use the same notation for the probability measure and its density with respect to the Lebesgue measure). Typically it is not possible to compute $\pi(f)$ analytically, and a common solution is to use approximations based on Monte Carlo methods. Given  independent identically distributed observations $X_1,\ldots,X_{n}$ from $\pi$, we might estimate $\pi(f)$ by $\pi_n(f):=n^{-1}\sum_{k=1}^{n}f(X_k)$. The variance of such estimate equals $\sigma^2(f)/n$  with $\sigma^2(f)$ being the variance of the integrand with respect to $\pi$. The first way to obtain a tighter estimate $\pi_n(f)$ is simply to increase the sample size $n$. Unfortunately, this solution might be prohibitively costly, especially when the dimension $d$ is large enough and sampling from $\pi$ is complicated. An alternative approach is to decrease $\sigma^2(f)$ by constructing a new Monte Carlo experiment with the same expectation as the original one, but with a lower variance. Such methods are known as variance
reduction techniques. Introduction to many of them can be found in \cite{rubinstein2016simulation, GobetBook, glasserman2013monte}.

%Recently one witnessed a revival of interest in efficient variance reduction methods  for Markov chains, mostly with applications to MCMC algorithms; see for example \cite{dellaportas2012control}, \cite{mira2013zero}, \cite{oates:girolami:chopin:2016}, \cite{south:mira:drovandi:2018}, \cite{brosse2018diffusion} and references therein.

One of the popular approaches to variance reduction is the control variates method (see \citep{south2021post} and the references therein). It aims at constructing a cheaply computable random variable $\zeta$ (control variate) with $\PE[\zeta] = 0$ and $\PE[\zeta^2] < \infty$, such that the variance of the random variable $f(X) + \zeta$ is small, where $X \sim \pi$. One of the main difficulties here is to construct a class of control variates \(\zeta\) satisfying $\PE[\zeta]=0$. The complexity of this problem essentially depends on the degree of our knowledge on $\pi$.
For example, if \(\pi\) is analytically known and satisfies some regularity conditions, one can apply the well-known technique of  polynomial interpolation to construct control variates enjoying  some optimality properties, see, for example \citep[Section~3.2]{dimov2008monte}. Alternatively, if an orthonormal system in \(\ltwo(\pi)\) is analytically available, one can build control variates \(\zeta\) as a linear combination of the corresponding basis functions, see \citep{GobetCV}. Furthermore, if \(\pi\) is known only up to a normalizing constant (which is often the case in Bayesian statistics), one can apply the recent approach of constructing control variates  depending only on the gradient \(\nabla \log \pi\)  using either a Schr\(\ddot{\text{o}}\)dinger-type Hamiltonian operator in  \citep{assaraf1999zero,mira2013zero}, or the Stein operator in \citep{brosse2018diffusion}.
In some situations \(\pi\) is not known analytically, but \(X\) can be represented as a function of  simple random variables with known distribution.
Such  situation arises, for example, in the case of functionals of  discretized diffusion processes. In this case a Wiener chaos-type decomposition can be used to construct control variates with nice theoretical properties, see \citep{belomestny2018stratified}.
Note that in order to compare different  variance reduction approaches, one has to analyze their complexity, that is, the number of numerical operations required to achieve a prescribed magnitude of the resulting variance.
\par
Unfortunately, it is not always possible to generate independent observations distributed according to $\pi$. To overcome this problem one might consider MCMC algorithms, where the exact samples from $\pi$ are replaced by \((X_p)_{p \geq 0},\) forming a Markov chain with a marginal distribution of $X_n$ converging to $\pi$ in a suitable metric as $n$ goes to infinity. It is still possible to apply the control variates method in a similar manner to the plain Monte Carlo case, yet the choice of the optimal control variate becomes much more involved. Due to significant correlations between the elements of the Markov chain, it might be not enough to minimize the marginal variances of \((X_p)_{p \geq 0}\) as it was in independent case. Instead one may choose the control variate by minimizing the corresponding asymptotic variance of the chain as it is suggested in~\citet{belomestny2019esvm}. At the same time it is possible to express the optimal control variate in terms of the solution of the Poisson equation for the corresponding Markov chain \((X_p)_{p \geq 0}\). As it was observed in~\citet{henderson1997variance,henderson2004}, for a time-homogeneous Markov chain \((X_p)_{p \geq 0}\) with a stationary distribution \(\pi\), the function $U_{G}(x) := G(x)-\PE[G(X_{1})|X_0 = x]$ has zero mean with respect to $\pi$ for an arbitrary real-valued function \(G: \rset^d \to \rset\), such that \(G  \in L^1(\pi)\). Hence, $U_{G}(x)$ is a valid control functional for a suitable choice of $G$, with the best $G$ given by a solution of the Poisson equation
\begin{equation}
\label{eq:definition-poisson}
\CPE{G(X_{1})}{X_0 = x}-G(x)=-f(x)+\pi(f)  \eqsp.
\end{equation}
For such $G$ we obtain $f(x) - U_G(x) = f(x) - f(x) + \pi(f) = \pi(f)$ leading to an ideal estimator with zero variance. Despite the fact that the Poisson equation involves the quantity of interest \(\pi(f)\)  and can not be  solved explicitly in most cases, this idea still can be used to construct some  approximations for the optimal zero-variance control variates. For example,  \cite{henderson1997variance} proposed to compute approximations to the solution of the Poisson equation for specific Markov chains with particular emphasis on models arising in stochastic network theory. In \cite{dellaportas2012control} and \cite{brosse2018diffusion}  series-type control variates are introduced and studied for reversible Markov chains. It is assumed in \cite{dellaportas2012control}  that the one-step conditional expectations  can be computed explicitly  for a set of basis functions. \cite{brosse2018diffusion} proposed another approach tailored to diffusion setting which does not require the computation of integrals of basis functions and only involves  applications of the underlying generator. For more information on diffusion based algorithms we refer reader to the recent works \citep{dalalyan2017theoretical,durmus:moulines:2017, MR2353037, MR3861816}. Another family of variance reduction techniques aims at constructing a parametric class of control variates with zero mean with respect to the ergodic measure $\pi$. A popular choice is Stein control variates (see \citet{belomestny2019esvm,mira2013zero,oates:girolami:chopin:2016,south:mira:drovandi:2018,south2021post} and references therein).
\par
In this paper we propose a generic variance reduction method for additive functionals of Markov chains. %Moreover, we do not need to assume stationarity or/and sampling under the invariant distribution \(\pi.\)
Compared to Stein control variates techniques, the knowledge of the stationary distribution is not required. The variance reduction method we propose thus applies not only to MCMC methods (for which the distribution $\pi$ is known), but also to the more general setting in which the stationary distribution is not analytically known; such examples arise, in particular, when one wishes to integrate according to the stationary distribution of an ergodic diffusion or to estimate the value function (or the gradient of the value function) in reinforcement learning algorithms.

Compared to \cite{dellaportas2012control}, our approach is not restricted to $\pi$-reversible Markov kernels. We provide a non-asymptotic analysis for the so-called normal noise model, which covers as a special example the Langevin dynamics. We also consider variance reduction in the problem of  estimating the expectation of functions under the unknown stationary distribution of ergodic diffusion process.

The paper is organized as follows.  In \Cref{sec:setup} we set up the problem and introduce some notations. In \Cref{seq:mart_repr}, we outline the construction of a novel martingale representation. In \Cref{sec:variance-reduction} we show how this martingale representation can be used to construct control variates. In \Cref{sec:normal_noise} we analyze performance of the proposed variance reduction algorithm in case of the Markov chain, driven by the normal noise (see \Cref{sec:normal_noise} for the precise definition). Finally, in \Cref{sec:numerics} we illustrate our findings on different numerical examples.

\section{Setup}\label{sec:setup}
Our aim is to numerically compute  expectations of the form
\[
\pi(f)=\int_{\rset^d} f(x)\pi(\rmd x),
\]
where \(f:\) \(\rset^d \rightarrow \mathbb{R}\) and \(\pi\) is a probability measure supported on \(\rset^d\) equipped with its Borel $\sigma$-field.
If  $d$ is large and \(\pi(f)\) can not be computed analytically, one can apply Monte Carlo methods. However, in many practical situations  direct sampling from \(\pi\) is impossible and this precludes the use of plain Monte Carlo methods in this case. One popular alternative to Monte Carlo  is Markov Chain Monte Carlo (MCMC) where one is looking for a discrete time  (possibly non-homogeneous) Markov chain   \((X^{x}_p)_{p \in \nset_0}\) such that \(\pi\) is its unique invariant measure. In this paper we study a class of MCMC algorithms with \((X^{x}_p)_{p \in \nset_0}\) satisfying  the following recurrence relation:
\begin{equation}
\label{eq:chain_gen}
X^{x}_{p}=\Phi_{p}(X^{x}_{p-1},\xi_{p}),\quad p=1,2,\ldots ,\quad X_{0}=x
\end{equation}
for some i.i.d. random vectors \(\xi_p\in \mathbb{R}^m\) with distribution \(P_{\xi}\)
and some Borel-measurable
functions $\Phi_{p}\colon \rset^d \times\mathbb{R}^{m}\to \rset^d.$
In fact, this is quite general class of Markov chains (see \citet[Theorem~1.3.6]{moulines2018})
and many well-known MCMC algorithms can be represented in the form \eqref{eq:chain_gen}.
Let us consider two popular examples.
\begin{exmp}[Metropolis-Adjusted Langevin Algorithm]
The Metropolis-Hastings algorithm
associated with a target density \(\pi\) requires to choose a
%sequence of
proposal transition density \(q\). %\((q_p)_{p\geq 1}\).
The Markov chain is constructed as follows:
\begin{enumerate}
\item Given the previous state $X^{x}_p$, we generate a proposal \(Y_{p+1}\sim q(\cdot|X^{x}_p)\)\;
\item Accept the proposal  $X^{x}_{p+1}=
Y_{p+1}$  with probability $\alpha(X^{x}_p,Y_{p+1})$ where
\[
\textstyle{\alpha(y,y')=\min\left\{1,\frac{\pi(y')}{\pi(y)}\frac{q(y|y')}{q(y'|y)}\right\}.}
\]
Otherwise, set $X^{x}_{p+1}=X^{x}_p$.
\end{enumerate}
This transition is reversible with respect to \(\pi\) and therefore preserves the stationary density \(\pi\); see \cite[Chapter~2]{moulines2018}. If %\((q_p)_{p \geq 1}\)
$q$ has a wide enough support to eventually reach any region
of the state space  with positive mass
under \(\pi\), then this transition is irreducible and $\pi$ is a maximal irreducibility measure \cite{mengersen:tweedie:1996}. The  Metropolis-Adjusted Langevin algorithm (MALA) takes  \eqref{eq:chain} as proposal, that is,
\begin{eqnarray*}
q(y|x)=(\gamma)^{-d/2}\boldsymbol{\varphi}\Bigl([y-x+\gamma\mu(x)]/\sqrt{\gamma}\Bigr) \,.
\end{eqnarray*}
with $\boldsymbol{\varphi}(z) := (\sqrt{2\pi})^{-1}\rme^{-\|z\|^2/2}$ is a density of the standard normal random variable. It is not difficult to see that the MALA chain can be compactly represented in the form
\begin{align*}
X^{x}_{p+1} &=X^{x}_p+\mathbbm{1}\bigl(U_{p+1}\leq \alpha(X^{x}_{p},Y_{p+1})\bigr)(Y_{p+1}-X^{x}_p),  \\
Y_{p+1}&=X^{x}_p-\gamma\mu(X^{x}_p)+\sqrt{\gamma}Z_{p+1},
\end{align*}
where \((U_{p})_{p\geq 1}\) is an i.i.d. sequence of uniformly distributed on \([0,1]\) random variables independent of \((Z_p)_{p\geq 1}.\) Thus, we recover \eqref{eq:chain_gen} with  \(\xi_p=(U_p,Z_p)\in \mathbb{R}^{d+1}\) and
\begin{eqnarray*}
\Phi_p(x,(u,z)^\top)=x+\mathbbm{1}\bigl(u\leq \alpha(x,x-\gamma\mu(x)+\sqrt{\gamma}z)\bigr)(-\gamma\mu(x)+\sqrt{\gamma}z).
\end{eqnarray*}
\end{exmp}
\begin{exmp}
\label{ex:discr_diffusion}
  Let \((\X^{x}_t)_{t\geq 0}\) be the unique strong solution to SDE of the form:
\begin{eqnarray}
\label{eq:sde-inv}
\rmd \X^{x}_t=b(\X^x_t)\, \rmd t+\sigma(\X^x_t)\rmd W_t, \, \X_0 = x, \, t\geq 0,	
\end{eqnarray}
where \(b:\) \(\mathbb{R}^d\to \mathbb{R}^d\) and \(\sigma:\) \(\mathbb{R}^d\times \mathbb{R}^m \to \mathbb{R}^d\) are locally Lipschitz continuous functions with at most linear growth.
The process \((\X^x_t)_{t\geq 0}\) is a Markov
process and let \(L\) denote its infinitesimal generator defined by
\begin{equation*}
Lg=b^\top \nabla g+\frac{1}{2}\sigma^\top D^2g\sigma
\end{equation*}
for any \(g\in C_{0}^2(\rset^d).\)
If there exists a twice continuously differentiable Lyapunov function \(V:\) \(\mathbb{R}^d\to \mathbb{R}_{+}\) such that
\begin{eqnarray*}
\sup_{x\in \mathbb{R}^d} LV(x) <\infty,\quad \limsup_{|x|\to \infty} LV(x)<0,
\end{eqnarray*}
then there is an invariant probability measure \(\pi\)  
%for \(X,\) that is, \(X_t\sim \pi\) for all \(t> 0\) if \(X_0\sim \pi.\)
Invariant measures are crucial in the study of the long term behaviour of stochastic differential
systems \eqref{eq:sde-inv}.    Under some additional assumptions,  the invariant
measure \(\pi\) is  ergodic and this property
can be exploited  to compute  the integrals \(\pi(f)\) for \(f\in L^2(\pi)\) by means of ergodic averages. The idea is to replace the diffusion \(X\) by a (simulable) discretization scheme of the form (see e.g. \citep{MR3861816}, \citep{lamberton:pages:2002})
\begin{equation}
\label{eq:diffusion_discr}
X^{x}_{n+1}= X^{x}_n + \gamma_{n+1} b(X^{x}_n) + \sigma(X^{x}_n)(W_{\Gamma_{n+1}}-W_{\Gamma_n}), \quad n\geq 0,\quad X_0= x,
\end{equation}
where \(\Gamma_n=\gamma_1+\ldots+\gamma_n\) and \((\gamma_n)_{n\geq 1}\) is a non-increasing sequence of time steps. Then for a function \(f\in L^2(\pi)\) we can approximate \(\pi(f)\) via
\begin{eqnarray*}
\pi_n^\gamma(f)=\frac{1}{\Gamma_n}\sum_{i=1}^{n}\gamma_{i}f(X^{x}_i).
\end{eqnarray*}
Due to typically high correlation between \(X^{x}_1,X^{x}_2,\ldots\), variance reduction is of crucial importance here.
As a matter of fact, in many cases there is no explicit formula for the invariant measure and this makes the use of the Stein control functions (see e.g. \citep{mira2013zero, oates2017control}) impossible in this case. 
\par 
If $b = -\nabla U/2$ for some continuously differentiable function $U$, and $\sigma = 1$, the Markov chain~\eqref{eq:diffusion_discr} can be used to approximately sample from the density
\begin{equation}\label{eq:stationary_distr}
\pi(x)= Z^{-1} \rme^{-U(x)/2}, \quad Z= \int_{\mathbb{R}^{d}} \rme^{-U(x)/2}\, \rmd x,
\end{equation}
provided that \(Z < \infty\).
This method is usually referred to as Unadjusted Langevin Algorithm (ULA). In practice, a constant step-size discretization 
\begin{equation}
\label{eq:chain}
X^{x}_{p+1}=X^{x}_{p}-\gamma \nabla U (X^{x}_{p})/2 + \sqrt{\gamma}Z_{p+1},\, X^x_0 = x,
\end{equation}
is often considered, where $\left(Z_{p}\right)_{p\geq1}$ is an i.i.d. sequence of $d$-dimensional
standard Gaussian random vectors. Note that the invariant distribution \(\pi_\gamma\) of the chain
\eqref{eq:chain} is in general different from \(\pi\) and is not available analytically, although $\pi_{\gamma}$ converges to $\pi$ when $\gamma \rightarrow 0$, see  \citet{mattingly:stuart:higham:2002}, \citet{durmus:moulines:2017}. Hence the methods based on the Stein control variates will introduce additional bias when applied to \eqref{eq:chain}.
\end{exmp}

\section{Martingale representation}
\label{seq:mart_repr}
In this section we provide a general discrete-time martingale representation for  Markov chains of  type  \eqref{eq:chain_gen} which is  used later to construct an efficient variance reduction algorithm. Let \((\phi_k)_{k\in \mathbb{Z}_+}\) be a complete orthonormal system in \(\ltwo(\mathbb{R}^m, P_{\xi})\) with \(\phi_0\equiv 1\).  In particular, we have
\begin{equation*}
\mathsf{E}[\phi_i(\xi)\phi_j(\xi)]=\delta_{ij},\quad i,j\in  \mathbb{Z}_{+}
\end{equation*}
with \(\xi \sim P_{\xi}.\)
Notice that this implies that the random variables
$\phi_k(\xi)$, $k\ge1$, are centered. As an example, we can take  multivariate Hermite polynomials for the ULA algorithm and a tensor product of shifted Legendre polynomials for "uniform part" and Hermite polynomials for "Gaussian part"  of the random variable $\xi = (u, z)^T$ in MALA, as the shifted Legendre polynomials are orthogonal with respect to the Lebesgue measure on \([0,1].\)
\par
Let  $(\xi_p)_{p \in \nset}$ be i.i.d. $m-$dimensional random vectors with distribution $\mathsf{P}_{\xi}$. We denote via $(\mathcal{G}_p)_{p \in \nset_0}$  the filtration generated by $(\xi_p)_{p \in \nset}$ with the convention $\mathcal{G}_0=\mathrm{triv}$.  Let $\Phi_k: \rset^d \times \rset^m \to \rset^d$ be a measurable function. Set for $l \le p$ and $x \in \rset^d$,
\begin{equation}
\label{eq:ula_new}
X^x_{l,p}:=G_{l,p}(x,\xi_{l+1},\ldots,\xi_{p})
\end{equation}
with the functions \(G_{l,p}:\) \(\rset^{d + m\times(p-l+1)}\to \rset^{d}\) defined as
\begin{equation}
\label{eq:definition-G-p-l}
G_{l,p}(x,y_{l+1},\ldots,y_p):=\Phi_p(\cdot,y_{p})\circ\Phi_{p-1}(\cdot,y_{p-1})\circ\dots\circ\Phi_{l+1}(x,y_{l+1})\,
\end{equation}
with the convention $G_{l,l}(x)= x$.
%Note that $\left(X^x_{0,p}\right)_{p \in \nset_0}$ is a Markov chain with values in $\rset^d$ of the form \eqref{eq:chain_gen}, starting at $X_0 = x$.
Note that for any bounded measurable function $f$, any $x \in \rset^d$ and $l \leq p,\, l,p \in \mathbb{N}$, it holds
\[
\CPE{f(X_p^x)}{\mcg_{l}} = \int [f\circ G_{l,p}](X_{l}^{x},e_{l+1},\ldots,e_p)\,P_{\xi}(\rmd e_{l+1})\ldots P_{\xi}(\rmd e_{p}) \eqsp.
\]
We write $X^x_{p}$ and $G_{p}$ as a shorthand notation for $X^x_{0,p}$ and $G_{0,p}$, respectively. We formulate the results below for bounded measurable functions, but these results can be easily extended to unbounded functions at the expense of  using classical drift conditions to control the moments. For simplicity and readability, we leave this elementary extension to the reader.
\begin{theorem}\label{thm:main-repr}
For any $q \in \mathbb{N}$, any $j < q, j \in \mathbb{N}$, any Borel bounded functions $f: \rset^d \rightarrow \rset$ and  $x \in \rset^d$ the following representation holds in \(\ltwo (\rset^{mq},P^{\otimes q}_{\xi})\)
\begin{equation}
\label{eq:mart_repr}
%f(X^x_{j, q}) = f(x) +\sum_{k=1}^{\infty}\sum_{l=j+1}^{q}a_{q,l,k}(X^x_{j,l-1})\phi_k\left(\xi_{l}\right),
f(X^x_{q}) = \mathsf{E}\left[\left.f(X^x_{q})\right|\mathcal G_{j}\right] +\sum_{k=1}^{\infty}\sum_{l=j+1}^{q}a_{q,l,k}(X^x_{l-1})\phi_k\left(\xi_{l}\right)
\end{equation}
where $X^x_{q}$ is given by \eqref{eq:ula_new} and for any $y \in \rset^d$,
\begin{equation}
\label{eq:coeff_mart}
%a_{p,l,k}(x)=\mathsf{E}\left[\left.f(X_{p})\phi_k\left(\xi_{l}\right)\right|X_{l-1}=x\right], \quad p\geq l, \quad k\in \mathbb{N}.
a_{q,l,k}(y)=\mathsf{E}\left[f(X^y_{l-1,q})\phi_k\left(\xi_{l}\right)\right], \quad q\geq l, \quad k\in \nset.
\end{equation}
\end{theorem}
\begin{proof}
The proof is postponed to Section~\ref{sec:proof:thm:main-repr}.
\end{proof}
\begin{corollary}
\label{coro:expansion:stationary}
Assume that $\Phi_l = \Phi$, for all $l\geq 1$. Then for any $q \in \nset$,  $j < q$, $f$ a bounded measurable function, and $x \in \rset^d$, it holds in \(\ltwo\bigl(\rset^{mq},P^{\otimes q}_{\xi}\bigr)\)
\begin{eqnarray*}
f(X^x_{q}) = \mathsf{E}\left[\left.f(X^x_{q})\right|\mathcal G_{j}\right] + \sum_{k=1}^{\infty}\sum_{l=j+1}^{q}\bar a_{q-l+1,k}(X^x_{l-1})\phi_k\left(\xi_{l}\right)
\end{eqnarray*}
where for all $y \in \rset^d$,
\begin{equation}
\label{eq:definition-bar-a-0}
\bar a_{r,k}(y)=\mathsf{E}\left[f(X^y_{r})\phi_k\left(\xi_{1}\right)\right] \quad  r, k\in \mathbb{N}.
\end{equation}
\end{corollary}

\paragraph{Discussion}
The representation \eqref{eq:mart_repr} is remarkable for two reasons. First, it suggests a general way of constructing zero-mean random variables adapted to the filtration \((\mathcal G_{p})_{p\geq 0}.\) Indeed any random variable of the form
\begin{eqnarray*}
\sum_{k=1}^{\infty}\sum_{l=j+1}^{q}\beta_{q,l,k}(X^x_{l-1})\phi_k\left(\xi_{l}\right)
\end{eqnarray*}
for some measurable functions \((\beta_{q,l,k})\) has zero mean (conditional on \(\mathcal G_{j}\)) and is adapted to \(\mathcal G_{q-1}.\) Second, it shows that  for coefficients defined in \eqref{eq:coeff_mart} the representation \eqref{eq:mart_repr} computes  exactly \(f(X^x_{q}), \) that is, the control variate
\begin{eqnarray*}
\sum_{k=1}^{\infty}\sum_{l=j+1}^{q}a_{q,l,k}(X^x_{l-1})\phi_k\left(\xi_{l}\right)
\end{eqnarray*}
is perfect and leads to zero variance when computing \(\CPE{f(X^x_{q})}{\mathcal G_{j}}\) by Monte Carlo.
Another equivalent representation of the coefficients \(a_{p,l,k}\)  turns out to be more useful in practice.
\begin{proposition}
Let $q\geq l, k\in \nset$. Then the coefficients \(a_{q,l,k}\) in \eqref{eq:coeff_mart}  can be alternatively represented as
\begin{equation}
\label{eq:definition-a-q-l-k}
a_{q,l,k}(x)=\mathsf{E}\left[\phi_k\left(\xi\right)Q_{l,q}\left(\Phi_l(x,\xi)\right)\right]
\end{equation}
with \(Q_{l,q}(y)=\mathsf{E}\left[f(X^y_{l,q})\right],\) \(q\geq l.\)
In the homogeneous case $\Phi_l=\Phi$, the coefficients $\bar{a}_{r,k}$ in \eqref{eq:definition-bar-a-0} are given respectively  for all $r \in \nset,$ by
\begin{equation}
\label{eq:definition-bar-a}
\bar a_{r,k}(x)=\mathsf{E}\left[\phi_k\left(\xi\right)Q_{r-1}\left(\Phi(x,\xi)\right)\right] \,
\text{ with }  Q_{r}(y)=\PE\left[f(X^y_r)\right], \quad r\in \mathbb{N}.
\end{equation}
\end{proposition}
We now show how the representation \eqref{eq:mart_repr} can be used to construct variable for of additive functionals of Markov chains. For the sake of clarity, in the sequel, we consider only the time homogeneous case ($\Phi_l = \Phi$ for all $l \in \nset$). For $f$ a bounded measurable function, denote
$\pi^x_{n}(f)= n^{-1} \sum_{p=1}^{n} f(X^x_{p})$, where $n\in\mathbb{N}$ is the number of samples.
To avoid overloading the notations, the dependence in the initial condition $x$ is removed when it can be inferred from the context
For any $q \in \mathbb{N}$, $k \in \mathbb{N}$, and $y \in \rset^d$, set
\begin{eqnarray}
\label{eq:definition-A-s,k}
A_{q,k}(y)=\sum_{r=1}^q \bar{a}_{r,k}(y).
\end{eqnarray}
\Cref{coro:expansion:stationary} applied with $j =0$ implies that for any $x \in \rset^d$,
\begin{equation}
\label{eq:expansion-estimator}
\pi^x_n(f)= \frac{1}{n} \sum_{q=1}^n \PE[f(X_q^x)] + \frac{1}{n} \sum_{k=1}^\infty
M^x_{n,k} \eqsp,  \text{with} \quad M^x_{n,k}= \sum_{l=1}^n A_{n-l+1,k}(X_{l-1}^x) \phi_k(\xi_l) \eqsp.
\end{equation}
Since \(\xi_{l}\) is independent of $\mathcal{G}_{l-1}$,  \(X^x_{l-1}\) is $\mathcal{G}_{l-1}$ measurable, and \(\mathsf{E}[\phi_k(\xi_{l})]=0,\) \(k\neq 0\), we get for  any measurable function $g$ with $\mathsf{E}\left[g^2(X_{l-1}^{x})\right] < \infty$ that
$\PE[g(X^x_{l-1})\phi_{k}\left(\xi_{l}\right)] =$ $\mathsf{E}[g(X^x_{l-1})\CPE{\phi_{k}\left(\xi_{l}\right)}{\mathcal{G}_{l-1}}] = 0$.
This implies that for any $k = 1,\dots,K$, $\bigl(M^x_{p,k}\bigr)_{p=1}^{\infty}$ is a square-integrable martingale sequence with respect to filtration $(\mathcal{G}_p)_{p \geq 1}$ and hence that, for any $n,k \in \nset$ and $x \in \rset^d$, $\PE[M^x_{n,k}]=0$. In addition, since $\PE[ \phi_k(\xi_l) \phi_{k'}(\xi_l)]=0$ if $k \ne k'$, we obtain for any $ 1 \leq k < k'$,
\begin{equation}
\label{eq:key-equation-variance-covariance}
\PVar(M^x_{n,k})= \sum_{\l=1}^n \PE[A^2_{n-l+1,k}(X_{l-1}^x)] \quad \text{and} \quad
\PCov(M^x_{n,k},M^x_{n,k'})= 0 \eqsp.
\end{equation}
Fix some $K\in\mathbb N$ and denote
\begin{equation}
\label{eq:M_K_n}
M^{(x,K)}_{n} = \sum_{k=1}^{K} M^x_{n,k} \eqsp.
\end{equation}
The expansion \eqref{eq:expansion-estimator} suggests to consider the following estimator
\begin{equation}
\label{eq:29032018a3}
\pi^{(x,K)}_{n}(f)=\pi^x_n(f)-n^{-1} M^{(x,K)}_{n}
\end{equation}
By construction, for any $n \in \nset$ and $x \in \rset^d$, \(\PE[\pi^{(x,K)}_{n}(f)]= \PE[\pi^x_n(f)]\) as \(\PE [M^{(x,K)}_{n}]=0.\)
Moreover,  we obtain
\begin{equation}
\label{eq:expression-variance}
\PVar[\pi^{(x,K)}_{n}(f)]=\frac{1}{n^2}\sum_{k=K+1}^{\infty} \PVar[M^x_{n,k}] \leq  \frac{1}{n^2}\sum_{k=1}^{\infty} \PVar[M^x_{n,k}]=\PVar[\pi^x_n(f)] \eqsp.
\end{equation}
Hence we expect $\PVar[\pi^{(x,K)}_{n}(f)]$ to be small, provided that $\PVar[M^x_{n,k}]$ decay fast enough as $k\to \infty$.

If the empirical mean estimator $\pi^x_n(f)$ is convergent in quadratic mean, the same is true for $\pi_n^{(x,K)}(f)$.
This is formalized in the following result. Denote by $P$ the Markov kernel of the Markov chain \eqref{eq:chain_gen}, defined for any bounded measurable function $f$ by  $P f(x)= \int f \circ \Phi(x,e) P_\xi(\rmd e)$.
\begin{proposition}
\label{prop:asymptotic-convergence}
Assume that the Markov kernel $P$ has a unique invariant probability measure $\pi$ and that for any bounded measurable function $f$, and $x \in \rset^d$,
\begin{equation}
\label{eq:convergence-condition}
\lim_{n \to \infty} \frac{1}{n} \left( \sum_{p=1}^n P^p f(x) - \pi(f) \right)^2 = 0 \quad \text{and} \quad
\lim_{n \to \infty} n \PVar[ \pi_{n}^x(f)] = \sigma_\pi^2(f) \eqsp.
\end{equation}
Then, for any $K \in \nset$,
\[
\limsup_{n \to \infty} n \PE[ \{ \pi_n^{(x,K)}(f) - \pi(f) \}^2] \leq \lim_{n \to \infty} n \PVar[ \pi_{n}^x(f)] = \sigma_\pi^2(f) \eqsp.
\]
\end{proposition}
The proof of \Cref{prop:asymptotic-convergence} is an elementary consequence of \eqref{eq:expression-variance} and is left to the reader.
A direct consequence is that if the sequence of estimator $\{ \pi_n^x(f) \}_{n=1}^\infty$ is consistent in quadratic mean then $\{ \pi_n^{(x,K)}(f) \}_{n=1}^\infty$ is also consistent in quadratic mean. For any $n \in \nset$ and $x \in \rset^d$, $\PE[ \pi_n^x(f)]= \PE[ \pi_n^{(x,K)}(f) ]$ and the variance of $\pi_n^{(x,K)}(f)$ is always smaller than that of $\pi_n^x(f)$.

Below is a simple  illustrative example showing that $\PVar[\pi^{(x,K)}_{n}(f)]$ can be much smaller than $\PVar[\pi^x_{n}(f)]$ even for $K=1$.
\begin{exmp}
\label{ex:full_decomposition}
Suppose that we aim at sampling from the Gaussian distribution with zero mean and variance $1/2$ with the density
$\pi(x) = (\sqrt{\pi})^{-1}\rme^{-x^2}\,$
using the ULA algorithm (see \Cref{ex:discr_diffusion} and equation \eqref{eq:chain}). We consider the Markov chain given by
%\[
%\theta_{p}=\bigl(1-\frac{\gamma}{\sigma^2}\bigr)\theta_{p-1} + \sqrt{\gamma}\xi_{p},\quad p\in\mathbb{N},
%\]
\begin{equation}
\label{eq:ULA_chain_gaussian}
X_{p}^{x}=\bigl(1-\gamma\bigr)X_{p-1}^{x} + \sqrt{\gamma}\xi_{p}, \quad X_{0}^x = x, \quad p\in\mathbb{N}, \quad \gamma \in (0,1);
\end{equation}
where $(\xi_{p})_{p \geq 1}$ is an i.i.d. sequence of normally distributed random variables with zero mean and unit variance. The invariant distribution of this Markov chain is Gaussian with zero mean and variance $1/(2-\gamma)$. As a complete orthogonal system in $\ltwo(\mathbb{R}, P_{\xi})$, we consider the normalized Hermite polynomials on $\mathbb{R}$, that is,
\begin{eqnarray}
\label{eq:herm1d}
H_k(x):=\frac{(-1)^k}{\sqrt{k!}}\rme^{x^2/2}\frac{\partial^k}{\partial x^k}\rme^{-x^2/2},
\quad x \in \mathbb{R}, \quad k \in \nset.
\end{eqnarray}
Consider now the problem of estimating $\pi(f)$ for $f(x)=x^{2}$. Note that
\begin{equation*}
f(X_{p}^x) - \mathsf{E}\left[f(X_{p}^x)\right] = \bigl(1-\gamma\bigr)^{2}\bigl(f(X^{x}_{p-1}) - \PE\bigl[f\bigl(X^{x}_{p-1}\bigr)\bigr]\bigr)
+ 2\bigl(1-\gamma\bigr)X^{x}_{p-1}\sqrt{\gamma}\xi_{p} + \gamma\left(\xi_{p}^{2}-1\right)\,,
\end{equation*}
and by recalling the definition of the Hermite polynomials, we arrive at the martingale
representation
\begin{align}
\label{eq:fxp}
f(X_{p}^x) - \PE\bigl[f\bigl(X_{p}^x\bigr)\bigr] %=\sum_{l=1}^{p}\bigl(1-\frac{\gamma}{\sigma^2}\bigr)^{2(p-l)}\left[2\bigl(1-\frac{\gamma}{\sigma^2}\bigr)\theta_{l-1}\sqrt{\gamma}\xi_{l} + \gamma\left(\xi_{l}^{2}-1\right)\right]\\
=\sum_{l=1}^{p}\bar{a}_{p-l+1,1}(X^x_{l-1})H_{1}(\xi_{l})+\sum_{l=1}^{p}\bar{a}_{p-l+1,2}(X^x_{l-1})H_{2}(\xi_{l}),
\end{align}
where for $z \in \rset$ and $q \in \nset$,
\begin{equation}
\label{eq:bar-a-AR}
\bar{a}_{q,1}(z)=2\bigl(1-\gamma \bigr)^{2q-1}\sqrt{\gamma}z,\quad \bar{a}_{q,2}(z)=\bigl(1-\gamma\bigr)^{2(q-1)}\gamma
\end{equation}
%and $H_{1}(z)=z,$ $H_{2}(z)=z^{2}-1$ are Hermite polynomials.
We stress that the coefficients $\bar{a}_{q,2}$ do not depend on $z$ in this special example. Due to \eqref{eq:definition-A-s,k} and \eqref{eq:fxp}, we can represent $\pi^x_n(f)$ as
\begin{align}
\nonumber
\pi^x_n(f) &= n^{-1}\sum\limits_{p=1}^{n}\mathsf{E}\left[f(X_p^x)\right] + n^{-1}\sum_{l=1}^{n}A_{n-l+1,1}(X^x_{l-1})H_{1}(\xi_{l}) + n^{-1}\sum_{l=1}^{n}A_{n-l+1,2}(X^x_{l-1})H_{2}(\xi_{l}) \\
 &= n^{-1}\sum\limits_{p=1}^{n}\mathsf{E}\left[f(X_p^x)\right] + n^{-1}M^x_{n,1} + n^{-1}M^x_{n,2},
 \label{eq:pi_n_example_decomposition}
\end{align}
where, for any $z \in \rset$ and $q \in \nset$,
\[
A_{q,1}(z) %= 2\sum_{p=l}^{n}\bigl(1-\frac{\gamma}{\sigma^2}\bigr)^{2(p-l)+1}\sqrt{\gamma}z
= 2z\sqrt{\gamma}\bigl(1-\gamma\bigr)\frac{1-\bigl(1-\gamma\bigr)^{2(q+1)}}{1-\bigl(1-\gamma\bigr)^{2}}, \quad A_{q,2}(z)
=\gamma\frac{1-\bigl(1-\gamma\bigr)^{2(q+1)}}{1-\bigl(1-\gamma\bigr)^{2}}.
\]
%Denote $\pi_{n,1}(f)= \pi^x_n(f) - n^{-1}M_{n,1}$, and $\pi_{n,2}(f)= \pi_n(f) - n^{-1}M_{n,1} - n^{-1}M_{n,2}$.
The decomposition \eqref{eq:pi_n_example_decomposition} implies that
\begin{equation}
\label{eq:var_bound_pi_n_gaus}
\begin{split}
&\PVar\left(\pi^x_{n}(f)\right)
=  \frac{1}{n^{2}}\sum_{l=1}^{n}\mathsf{E}\left[A_{n-l+1,1}^{2}(X^x_{l-1})\right]+\frac{1}{n^{2}}\sum_{l=1}^{n}\mathsf{E}\left[A_{n-l+1,2}^{2}(X^x_{l-1})\right]\,, \\
&\PVar\bigl(\pi_{n}^{(x,1)}(f)\bigr)
= \frac{1}{n^{2}}\sum_{l=1}^{n}\mathsf{E}\left[A_{n-l+1,2}^{2}(X^x_{l-1})\right]\,.
\end{split}
\end{equation}
Hence, given that $\gamma < 1$, we estimate
%\begin{align*}
%\PVar\bigl(\pi^{(x,1)}_n(f)\bigr) \leq \frac{\gamma^2}{n\bigl(1-\bigl(1-\gamma\bigr)^{2}\bigr)^{2}} \leq \frac{1}{n},
%\end{align*}
%\begin{align*}
%\sum_{l=1}^{n}\mathsf{E}\left[A_{n-l,1}^{2}(X^x_{l-1})\right] & =\frac{4\gamma\bigl(1-\gamma\bigr)^{2}}{\bigl(1-\bigl(1-\gamma\bigr)^{2}\bigr)^{2}}\sum_{l=1}^{n}\PE\left[f(X^x_{l-1})\right]\bigl(1-\bigl(1-\gamma\bigr)^{2(n-l+1)}\bigr)^{2}
%\end{align*}
%\begin{align*}
%\sum_{l=1}^{n}\mathsf{E}\left[A_{n-l,2}^{2}(X^x_{l-1})\right] = \frac{\gamma^{2}}{\bigl(1-\bigl(1-\gamma\bigr)^{2}\bigr)^{2}}\sum_{l=1}^{n}\bigl(1-\bigl(1-\gamma\bigr)^{2(n-l+1)}\bigr)^{2} \leq \frac{n\gamma^2}{\bigl(1-\bigl(1-\gamma\bigr)^{2}\bigr)^{2}}
%\end{align*}
\begin{align*}
\mathsf{Var}\left(\pi^{(x,1)}_{n}(f)\right) & =
\frac{1}{n^2}\sum\limits_{l=1}^{n}\mathsf{E}\left[A_{n-l,2}^2(X^x_{l-1})\right] \leq \frac{\gamma^2}{n\bigl(1-\bigl(1-\gamma\bigr)^{2}\bigr)^{2}} \leq \frac{1}{n}\,,
\end{align*}
which does not depend upon $x$. On the other hand, $\PVar\left(\pi^x_{n}(f)\right) = n^{-2}\sum_{i,j}\PCov{\bigl(f(X_i^x),f(X_j^x)\bigr)}$, and since $X_i^x$
and $X_j^x$ are Gaussian random variables, application of the Isserlis formula yields
\begin{align*}
\PCov{\bigl(f(X_i^x),f(X_j^x)\bigr)} = 2\PCov^2{\bigl(X_i^x,X_j^x\bigr)} + 4\PCov{\bigl(X_i^x,X_j^x\big)}\mathsf{E}\bigl[X_i^x\bigr]\mathsf{E}\bigl[X_j^x\bigr].
\end{align*}
Using the identity $\PCov{\bigl(X_i^x,X_j^x\bigr)} = \gamma \sum_{k=1}^{i \wedge j}(1-\gamma)^{i+j-2k} \geq (1/2)\bigl((1-\gamma)^{|i-j|} - (1-\gamma)^{i+j}\bigr)$, we get
\begin{align*}
&\PVar\left(\pi^x_{n}(f)\right) \geq \frac{1}{4(n\gamma)} + \frac{x^2}{18(n\gamma)^2}, \quad  n\gamma > 2\,, \gamma \leq 1/2.
%&\frac{4\gamma \bigl(1-\gamma\bigr)^2}{n^2\left(1-\bigl(1-\gamma\bigr)^{2}\right)^{2}}\sum_{l=1}^{n}\mathsf{E}\left[f(X^x_{l-1})\right]\left(1-\bigl(1-\gamma\bigr)^{2(n-l+1)}\right)^{2} \geq \\
% & \geq \frac{4 \gamma \bigl(1 -\gamma \bigr)^2}{n^2\left(1-\bigl(1-\gamma \bigr)^{2}\right)^{2}}\sum\limits_{l=\lceil\frac{n}{4}\rceil}^{\lceil\frac{3n}{4}\rceil}\mathsf{E}\left[f(X^x_{l-1})\right]\bigl(1 - \rme^{-2\gamma(n-l+1)}\bigr)^2 \gtrsim
\end{align*}
Thus, for $\gamma \in (0,1/2]$, $\mathsf{Var}\left(\pi^{(x,1)}_{n}(f)\right) \Bigl / \mathsf{Var}\left(\pi^x_{n}(f)\right) \leq 4\gamma$, and the variance reduction effect is large when  $\gamma \downarrow 0^{+}$. To make the variance reduction effect clear, we plot the ratio $\PVar\left(\pi^x_{n}(f)\right)/\PVar\bigl(\pi_{n}^{(x,1)}(f)\bigr)$, computed according to \eqref{eq:var_bound_pi_n_gaus} with $n = 10^4$, $x = 1$ and different values of the step size $\gamma$. The corresponding plots are provided in \Cref{fig:vr_plot_gaus}. It illustrates first that the gain in variance indeed scales linearly in $\gamma$ for $\gamma$ small enough, and, second, that even for moderate values of $\gamma$, estimate $\pi^{(x,1)}_{n}(f)$ is preferable in terms of variance.
\qed
\end{exmp}

\begin{figure}[tbh]
\centering
\subfloat{
\includegraphics[width=0.45\textwidth]{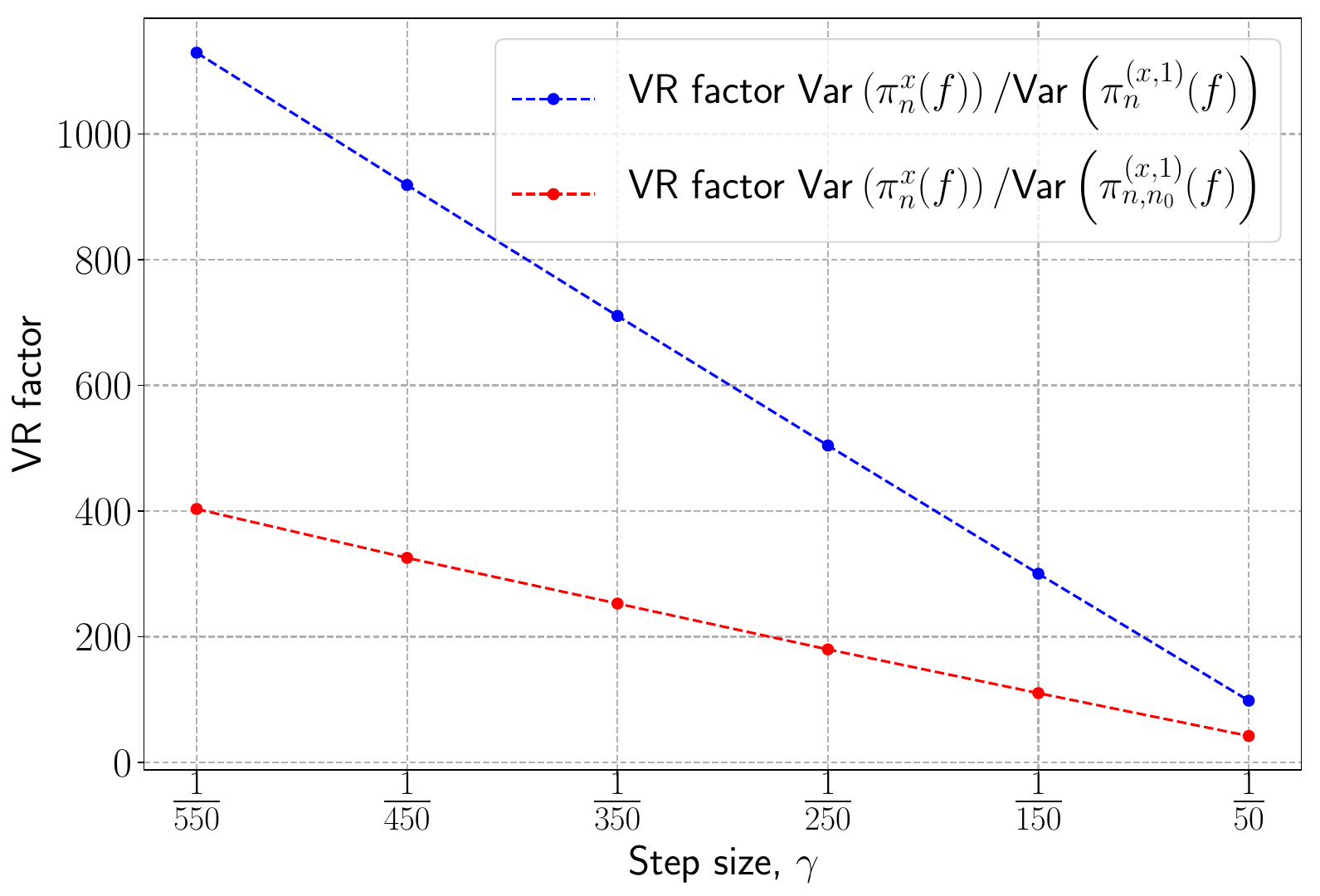}
\label{fig:subfig1}}
\qquad
\subfloat{
\includegraphics[width=0.45\textwidth]{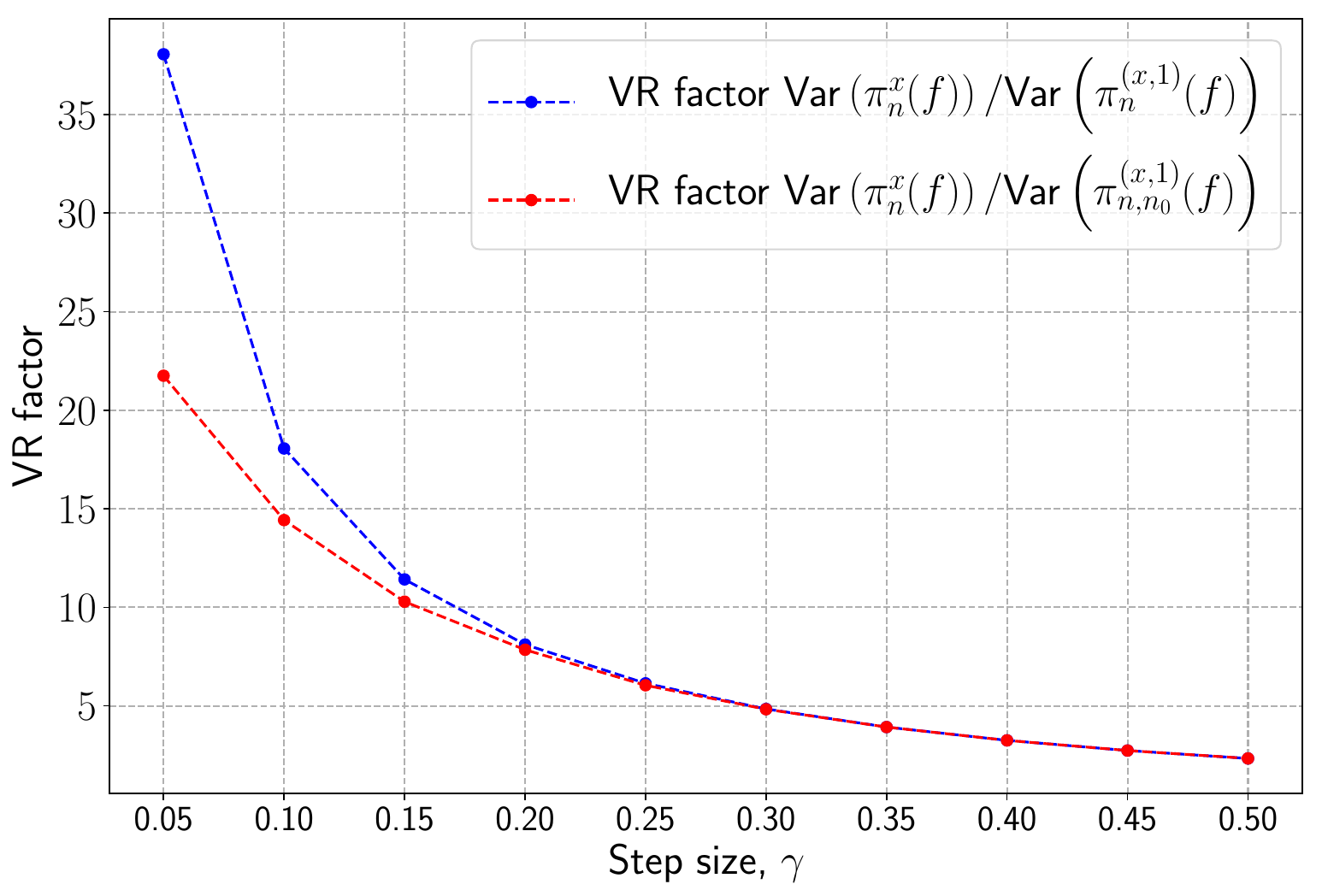}
\label{fig:subfig2}}
\caption{Variance reduction ratios for the estimates $\pi_{n}^{(x,1)}(f)$ (see \eqref{eq:29032018a3}) and $\pi^{(x,1)}_{n,\ntrunc}(f)$ (see \eqref{eq:VR-estimator}) for \Cref{ex:full_decomposition} as functions of step size $\gamma$ for different ranges of $\gamma$ values. We fix $n = 10^4$, $X_0 = x = 1$, and set truncation point $\ntrunc$ according to $\ntrunc = 5 + \lceil \log{\gamma^{-1}}/(4\gamma)\rceil$.
\label{fig:vr_plot_gaus}}
\end{figure}

\section{Martingale Decomposition Control Variate (MAD-CV) algorithm}
\label{sec:variance-reduction}
We now describe an algorithm to estimate the martingale $(M^{(x,K)}_{n})$ introduced in \eqref{eq:M_K_n}.
To keep the computational complexity at a reasonable level, we estimate a fixed number of coefficients $\bar{a}_{r,k}$ for $r = 1,\ldots,\ntrunc$, where the truncation index $\ntrunc$ does not depend on $n$. The corresponding martingale then is written as $M^{(x,K)}_{n,\ntrunc} = \sum_{k=1}^{K}M^x_{n,k,\ntrunc}$ where
\begin{equation}
\label{eq:M_K_n_truncated}
M^x_{n,k,\ntrunc} = \sum_{l=1}^{n}
A_{n-l+1,k,\ntrunc}(X_{l-1}^{x})\phi_{k}(\xi_{l}), \quad A_{q,k,\ntrunc}(y)=\sum_{r=1}^{q \wedge \ntrunc} \bar{a}_{r,k}(y),
\end{equation}
and we consider the truncated version of the estimator \eqref{eq:29032018a3}:
\begin{equation}
\label{eq:VR-estimator}
\pi^{(x,K)}_{n,\ntrunc}(f) = \pi^x_n(f) - n^{-1}M^{(x,K)}_{n,\ntrunc}.
\end{equation}
The truncation leads to an increase of the variance. More precisely, proceeding as in \eqref{eq:expression-variance}, we obtain
\begin{eqnarray}
\label{eq:expression-variance-trunc}
\nonumber
\PVar[\pi^{(x,K)}_{n,\ntrunc}(f)]&=&\frac{1}{n^2}\sum_{k=K+1}^{\infty} \PVar[M_{n,k}^x]
\\
&&+\frac{1}{n^2}\sum_{k=1}^{K}\sum_{l=1}^{n}\PE[(A_{n-l+1,k}(X^x_{l-1})-A_{n-l+1,k,\ntrunc}(X^x_{l-1}))^2] \,
\end{eqnarray}
where $A_{r,k}$ is defined in \eqref{eq:definition-A-s,k}. To illustrate the effect of truncation on the variance reduction, we compare $\PVar[\pi^{(x,K)}_{n,\ntrunc}(f)]$ and $\PVar[\pi^{(x,K)}_{n}(f)]$ using the simple \Cref{ex:full_decomposition}.
\begin{continueexample}{ex:full_decomposition}
We now consider the estimate $\pi^{(x,1)}_{n,\ntrunc}(f) = \pi^x_n(f) -n^{-1}M^x_{n,1,\ntrunc}$ where $M^x_{n,1,\ntrunc}$ is the truncated martingale
\[
M^x_{n,1,\ntrunc} := \sum\limits_{p=1}^{n}\sum\limits_{l=1}^{p \wedge \ntrunc } \bar{a}_{l,1}(X^x_{p-l-1})H_{1}(\xi_{p-l}) \, ,
\]
where $\bar{a}_{l,i}$, $i=1,2$ is defined in \eqref{eq:bar-a-AR}.
Then, proceeding as in \Cref{ex:full_decomposition}, we obtain
\begin{align*}
\PVar\left(\pi^x_n(f) -n^{-1}M^x_{n,1,\ntrunc}\right) %&= \frac{1}{n^{2}}\sum_{l=1}^{n}\mathsf{E}\left[A_{n-l,2}^{2}(X_{l-1})\right] + \frac{1}{n^2}\mathsf{Var}\left(\sum\limits_{p=\ntrunc+1}^{n}\sum\limits_{k=\ntrunc}^{p-1}a_{k,1}(X_{p-k-1})H_1(\xi_{p-k})\right) = \\
%&= \frac{1}{n^{2}}\sum_{l=1}^{n}\mathsf{E}\left[A_{n-l,2}^{2}(\theta_{l-1})\right] + \frac{1}{n^2}\mathsf{Var}\left(\sum\limits_{p=r+1}^{n}\sum\limits_{u=1}^{p-r}a_{p-u,1}(\theta_{u-1})H_1(\xi_{u})\right) = \\
%&= \frac{1}{n^{2}}\sum_{l=1}^{n}\mathsf{E}\left[A_{n-l,2}^{2}(\theta_{l-1})\right] + \frac{1}{n^2}\mathsf{Var}\left(\sum\limits_{d=1}^{n-r}\sum\limits_{u=1}^{d}a_{d+r-u,1}(\theta_{u-1})H_1(\xi_{u})\right) = \\
= \frac{1}{n^{2}}\PVar[M_{n,2}^x] +
\underbrace{\frac{1}{n^2}\sum\limits_{u=1}^{n-\ntrunc}\PE\Bigl[\sum_{l=u}^{n-\ntrunc}\bar{a}_{l+\ntrunc-u,1}(X^{x}_{u-1})\Bigr]^2}_{R_{\gamma,\ntrunc}} \, ,
\end{align*}
%Note that
%\begin{equation*}
%\sum\limits_{d=u}^{n-r}a_{d+r-u,1}(X_{u-1}) = 2\bigl(1-\gamma\bigr)^{2r+1}\sqrt{\gamma}X_{u-1}\frac{1-\bigl(1-\gamma\bigr)^{2(n-r-u+1)}}{1-\bigl(1-\gamma\bigr)^2}
%\end{equation*}
%Moreover, the recurrence
%\begin{equation*}
%\mathsf{E}\bigl[f(X_p^{x})\bigr] = \bigl(1-\gamma\bigr)^2 \mathsf{E}\bigl[f(X_{p-1}^x)\bigr] + \gamma
%\end{equation*}
%implies that
%\begin{equation*}
%\mathsf{E}\bigl[f(X^{x}_{p})\bigr] = \bigl(1 -\gamma \bigr)^{2p}x^2 + \gamma \frac{1 - (1-\gamma)^{2p}}{1 - \bigl(1-\gamma\bigr)^2}
%\end{equation*}
%Hence,
where $M_{n,2}^x$ is defined in \eqref{eq:pi_n_example_decomposition}.
Plugging \eqref{eq:bar-a-AR} in the previous identity, we get
\begin{equation*}
%\frac{1}{n^2}\mathsf{E}\sum\limits_{u=1}^{n-\ntrunc}\bigl[\sum\limits_{d=u}^{n-\ntrunc}a_{d+\ntrunc-u,1}(X^{x}_{u-1})\bigr]^2
R^x_{\gamma,\ntrunc} \leq \frac{4\gamma\bigl(1-\gamma \bigr)^{4\ntrunc+2}}{n^2\bigl(1 - \bigl(1-\gamma\bigr)^2\bigr)^{2}}\sum_{u=1}^{n-\ntrunc}\mathsf{E}\bigl[f(X_{u-1}^{x})\bigr]\bigl(1 - \bigl(1 - \gamma\bigr)^{2(n-\ntrunc-u+1)}\bigr)^{2}.
\end{equation*}
 Setting $\ntrunc = \lceil \log{\gamma^{-1}}/(4\gamma)\rceil$, we obtain $R^x_{\gamma,\ntrunc} \lesssim 1/n + x^2/(n^2\gamma)$ for $ n\gamma > 1$,  yielding the same (up to a constant factor) variance reduction factor, that is,
\[
\mathsf{Var}\left(\pi^{(x,1)}_{n,\ntrunc}(f)\right)\Bigl / \mathsf{Var}\left(\pi^x_{n}(f)\right) \lesssim \gamma,\quad n\gamma > 1.
\]
Here we write $\gtrsim$ and $\lesssim$ for inequality up to a constant not depending on $n, \gamma$ and $x$.
\end{continueexample}
The last step to define an estimator of the coefficients $A_{q,k,n_0}$. In the previous example, the calculation of these coefficients is explicit, but this is obviously not the case in general. We propose to use the representation outlined in \eqref{eq:definition-bar-a} of the
functions $\bar{a}_{r,k}$.
This representation suggests to first approximate the $r$-th step predictor $Q_{r}(y)=\PE\left[f(X^y_r)\right]$, $y \in \rset^d$,  for $r \in \{0,\dots,\ntrunc-1\}$.
For that purpose, we consider a parametric family of functions from $\rset^d$ to $\rset$, denoted $\{ \estQ{r}{\param}, \param \in \Param, r \in \{0,\dots,\ntrunc-1\}\}$,
where $\Param \subset \rset^{\qtrunc}$.
There are many ways to define such family of functions. The simplest idea  is to select a family of functions $\{\basis_b \}_{b=1}^{\qtrunc}$, $\basis_b: \rset^d \to \rset$ and to set
\begin{equation}
\label{eq:linear-estimator}
\estQ{r}{\param}(y) = \sum_{b=1}^{\qtrunc} \param_{b} \basis_{b}(y) \eqsp,
\end{equation}
However, this is not necessarily the best choice when the prediction functions $Q_r$ have a specific structure. 
For example, for Metropolis-Hastings algorithms, the sampling step uses an accept/reject step. In such case, it is more appropriate to consider predictors of the form
\begin{equation}
\label{eq:MCMC-linear-estimator}
\estQ{r}{\param}(y)= \estalpha{r}{\param}(y) f(y) + \testQ{r}{\param}(y) \eqsp.
\end{equation}
In this decomposition, $\estalpha{r}{\param}(y)$ estimates the $r$-th step rejection probability, \ie\ the probability of observing $r$ successive rejection. This probability can be estimated by logistic regression.

The parameter vector $\param$ are estimated via the least-squares approach. More precisely, for $r \in \{0,\dots,\ntrunc-1\}$, i.e. we solve
\begin{equation}
\label{eq:Qregr}
\hparam_r \in \argmin_{\beta \in \rset^\qtrunc} \sum_{s = 1}^{n-r}  \left| f(X^x_{r+s}) - \estQ{r}{\param}(X_s^x) \right|^2 \eqsp.
\end{equation}
Finally, we compute the estimates $\widehat a_{r,k}$ of the functions $\bar a_{r,k}$ (see \eqref{eq:definition-bar-a}). Namely, for all $y \in \rset^d$ we define
\begin{equation}
\label{eq:a-pol-gen}
\widehat a_{r+1,k} (y)
= \int \phi_{k}(z) \estQ{\hparam_r}{r}(\Phi(y,z)) P_{\xi}(\rmd z)
\end{equation}
where $\Phi$ is defined in \eqref{eq:chain_gen}. Note that in some relevant cases (e.g. for  \(\Phi\) being linear in \(z\) and the regression function being a linear combination of basis functions as in \eqref{eq:linear-estimator}), the expectation in \eqref{eq:a-pol-gen} can be computed in closed form. When direct integration is not an option, we  use Monte Carlo or Quasi Monte Carlo to compute the integrals $\int \phi_{k}(z) \basis_{b}(\Phi(y,z)) P_{\xi}(\rmd z)$. The complexity of this parametric integration problem is well studied. In order to increase efficiency, one can also employ the Multilevel Monte Carlo approach (see \citet{heinrich1999monte}). The estimator obtained by plugging \eqref{eq:a-pol-gen} into \eqref{eq:M_K_n_truncated}
and \eqref{eq:VR-estimator} is referred to as the MAD-CV (MArtingale Decomposition Control Variate) estimator.

The resulting estimate
\begin{equation}
\label{eq:vr_estimator_truncated}
\widehat\pi^{(x,K)}_{n,\ntrunc}(f) = \pi^x_n(f) - n^{-1}\widehat M^{(x,K)}_{n,\ntrunc}
\end{equation}
with
\begin{equation}
\label{eq:mad_cv_martingale_def}
\widehat M^{(x,K)}_{n,\ntrunc} = \sum_{k=1}^{K}\widehat M^{x}_{n,k,\ntrunc}, \quad
\widehat M^x_{n,k,\ntrunc} = \sum_{l=1}^{n}
\widehat A_{n-l+1,k,\ntrunc}(X_{l-1}^{x})\phi_{k}(\xi_{l}), \quad \widehat A_{q,k,\ntrunc}(y)=\sum_{r=1}^{q \wedge \ntrunc} \widehat a_{r,k}(y),
\end{equation}
remains unbiased for $\pi(f)$ (if computed on a new trajectory independent of regression data) and has a variance
\begin{eqnarray*}
\nonumber
\PVar[\widehat \pi^{(x,K)}_{n,\ntrunc}(f)]&=&\frac{1}{n^2}\sum_{k=K+1}^{\infty} \PVar[M_{n,k}^x]\\
&&+\frac{1}{n^2}\sum_{k=1}^{K}\sum_{l=1}^{n}\PE[(A_{n-l+1,k}(X^x_{l-1})-A_{n-l+1,k,\ntrunc}(X^x_{l-1}))^2]
\\
&&+\frac{1}{n^2}\sum_{k=1}^{K}\sum_{l=1}^{n}\PE[(\widehat A_{n-l+1,k,\ntrunc}(X^x_{l-1})-A_{n-l+1,k,\ntrunc}(X^x_{l-1}))^2].
\end{eqnarray*}
While the first and the second terms will be studied in \Cref{sec:normal_noise} in some special cases,  the last one has to be analyzed separately for different approximation schemes, and we leave this analysis for future research. Nevertheless already this decomposition shows that under the conditions of Proposition~\ref{prop:asymptotic-convergence}, it holds for a fixed $K>0,$
\begin{align*}
\limsup_{n\to\infty}n\mathsf{E}[\{\widehat{\pi}_{n,n_{0}}^{(x,K)}(f)-\pi(f)\}^{2}]&\leq\limsup_{n\to\infty}n\mathsf{E}[\{\pi_{n}^{x}(f)-\pi(f)\}^{2}]
\\
&+\limsup_{n\to\infty}n\mathsf{Var}[\widehat{\pi}_{n,n_{0}}^{(x,K)}(f)]\leq\widehat{\sigma}_{K}^{2}(f)<\infty,
\end{align*}
provided that  the expectations  \(\PE[A^2_{n-l+1,k}],\) \(\PE[\widehat A^2_{n-l+1,k,\ntrunc}]\) are uniformly bounded for $l=1,\ldots,n,$  in $n\in \mathbb{N}$ and $k=1,\ldots,K$. The latter property of the estimates $\widehat A_{n-l+1,k,\ntrunc}$ can be achieved by using an additional truncation step in regression, see e.g. \cite{gyorfi2006distribution} for various truncation schemes.

%Finally, we construct the truncated version of the estimator \eqref{eq:29032018a3}:
%\[
%\pi^{(K)}_{n,\ntrunc}(f) = \pi_n(f) - n^{-1}\widehat M^{(K)}_{n,\ntrunc}
%\]
%where
%\begin{equation}
%\label{eq:m_trunc}
%\widehat M^{(K)}_{n,\ntrunc} = \sum_{p=\ntrunc}^{n}\left[\sum_{k=1}^{K}\sum_{l=p-\ntrunc+1}^{p} \widehat a_{p-l,k}(X^{x}_{l-1})\phi_{k}(\xi_{l})\right].
%\end{equation}

\section{Gaussian noise model}
\label{sec:normal_noise}
We analyze the MAD-CV algorithm for the Markov chains $(X^{x}_p)_{p\geq 0}$ driven by a normal noise, that is,
\begin{equation}
\label{eq:chain_norm}
X^{x}_{p}=\Phi(X^{x}_{p-1}, Z_{p}),\quad Z_p\sim \mathcal{N}(0,\Id_{d\times d}),\quad p=1,2,\ldots , \quad X^{x}_{0}=x
\end{equation}
%We put a $\sqrt{\gamma}$ factor here, since  we are interested in the behavior of the variance reduced estimator \eqref{eq:29032018a3} in a ``small noise'' regime, that is,  in the case when $\gamma\to 0^{+}$.
For a multi-index $\kind=(k_i)\in \nset_0^d$, we denote by $\mathbf{H}_\kind(x)$ the normalized Hermite polynomial on $\rset^d$, that is,
$\mathbf{H}_\kind(x):=\prod_{i=1}^d H_{k_i}(x_i),\, x=(x_i) \in \rset^d$  with $H_{k_i}$ defined in \eqref{eq:herm1d}. The following notations are used in the sequel: $\|\kind\|=\max\limits_{i \in \{1,\ldots,d\}} k_i$, $|\kind|=\sum_{i=1}^d k_i$ and $\kind! :=k_1!\dots k_d!$. In this case $a_{r,\kind}(y)=\mathsf{E}\left[f(X^y_{r}) \mathbf{H}_\kind\left(Z_{1}\right)\right]$, $A_{q,\kind}(y)=\sum_{r=1}^q \bar{a}_{r,\kind}(y)$ and the martingale $M^{(x,K)}_{n}$ takes the form
\begin{equation}
\label{eq:multidim_decomposition}
M^{(x,K)}_{n} = \sum_{0 < \|\kind\| \leq K}\sum_{l=1}^{n}
A_{n-l+1,\kind}(X_{l-1}^{x}) \mathbf{H}_\kind(Z_{l}).
\end{equation}
Recall that $\mathcal G_p=\sigma(Z_1,\ldots,Z_p)$, $p\in\mathbb N$, and $\mathcal G_0=\mathrm{triv}$. For a twice differentiable function $g: \rset^d \rightarrow \rset$, we denote by $D^2g(x)$ its Hessian at point $x$. For a smooth function $g\colon\mathbb \rset^d \to \rset$, a multi-index $\kind \in \nset_0^d$, we use the notation $g^{(\kind)}(x)$ for the partial derivative
\[
g^{(\kind)}(x)
:= \partial^{k_1}_{x_1}
\ldots
\partial^{k_d}_{x_d}
g(x),
\quad x = (x_1,\ldots,x_d),
\]
For $m\in\mathbb N$, a smooth function
$h\colon\mathbb \rset^{d\times m}\to \rset$
with arguments being denoted
$(z_1,\ldots,z_m)$, $z_i\in\mathbb R^d$, $i=1,\ldots,m$,
a multi-index $\mathbf k=(k_i)\in\mathbb N_0^d$,
and $j\in\{1,\ldots,m\}$,
we use the notation $\partial^{\mathbf k}_{z_j} h$ for the multiple derivative of $h$
with respect to the components of~$z_j$:
\[
\partial^{\mathbf k}_{z_j} h(\chunk{z}{1}{m})
:=\partial^{k_1}_{z_{j,1}}
\ldots
\partial^{k_d}_{z_{j,d}}
h(z_1,\ldots,z_m),
\quad z_j = \bigl( z_{j,1},\ldots,z_{j,d} \bigr),
\]
where  $\chunk{z}{1}{m}= (z_1,\dots,z_m)$ and
\(\partial^{k_s}_{z_{j,s}} h(\chunk{z}{1}{m}) \) stands for partial derivative of order \(k_s\) for the function \(h\) with respect to the \(s\)th - coordinate of the vector \(z_j.\) For $m \in \nset$ and $j \leq m$, we denote
\begin{equation}
\label{eq:definition-gradient}
\nabla_{z_j}h(\chunk{z}{1}{m}) = \bigl(\partial_{z_j^1}h(\chunk{z}{1}{m}), \dots, \partial_{z_j^d}h(\chunk{z}{1}{m})\bigr)
\end{equation}
By setting $G_p=G_{0,p}$, where $G_{0,p}$ is defined in \eqref{eq:definition-G-p-l}, we obtain
$f\left(X_{p}^x\right) = [f\circ G_{p}](x,\chunk{Z}{1}{p})$.

We preface the derivations with an auxiliary lemma which provides us with a representation for the coefficients $\bar a_{p,\kind}$ defined in \eqref{eq:definition-bar-a}. Let $K \in \nset$ and consider the following assumptions
\begin{assumption}
\label{assum:regularity_Phi}
The function $\Phi: \rset^d \times \rset^m \to \rset^d$ is $K \times d$ times continuously differentiable.
\end{assumption}
\begin{assumption}
\label{assum:regularity_f}
The function $f: \rset^d \to \rset$ is $K \times d$ times continuously differentiable.
\end{assumption}

\begin{lemma}
\label{lem:a_repr}
Assume \Cref{assum:regularity_Phi} and \Cref{assum:regularity_f} and that \eqref{eq:chain_norm} holds. Then for any \(\kind,\kind^{\prime} \in \mathbb{N}_0^d\) such that \(\kind '\le \kind\) componentwise and $\| \kind' \| \leq K$, any $x \in \rset^d$, the following representation holds
\[
\bar a_{p,\kind}(x) = \left(\frac{(\kind-\kind') !}{\kind!}\right)^{1/2}
\PE\left[
\partial_{z_1}^{\kind'}[f\circ G_{p}](x,\chunk{Z}{1}{p})\mathbf{H}_{\kind-\kind'}(Z_1)\right].
\]
\end{lemma}
\begin{proof}
The proof is postponed to \Cref{proof:lem:a_repr}.
\end{proof}

Under some additional smoothness assumptions one can derive a useful bound for the sum of the functions $A^2_{q,\kind}$ which can be directly used to bound the variance of additive functionals \eqref{eq:expression-variance}.
\begin{proposition}
\label{prop:main}
Assume \Cref{assum:regularity_Phi}, \Cref{assum:regularity_f}, and that \eqref{eq:chain_norm} holds.
%Suppose that \eqref{eq:chain_norm} holds for a function $\Phi$, such that the derivatives
%\[
%D^{I}_j(x,y_1,\ldots,y_p)=\partial_{y_{j}}\partial_{y_{1}}^{\bigK_{I}} [f\circ G_{p}](x,y_{1},\ldots,y_p),\quad j=1,\ldots, p,\quad p\in \mathbb{N}
%\]
Then for any $x \in \rset^d$, it holds
%\begin{eqnarray*}
%\sum_{\kind \colon\| \kind \|\geq K+1}A^2_{q,\kind}(x)
%\leq \sum_{I\subseteq\{1,\ldots,d\},\, I\neq \emptyset}
%\left(\frac{\gamma}{2}\right)^{|I|K}
%\sum_{j=1}^{q}\PE\biggl[\bigl(\gamma\sum_{p=j}^{q }D^{I}_j(x,\sqrt{\gamma}Z_1,\ldots, \sqrt{\gamma}Z_p)\bigr)^{2}\biggr].
%\end{eqnarray*}
\begin{equation}
\label{eq:sumA_lemma}
\sum_{\kind \colon\| \kind \|\geq K+1}A^2_{q,\kind}(x)
\leq \sum_{I\subseteq\{1,\ldots,d\},\, I\neq \emptyset}
(1/2)^{|I|K}
\sum_{j=1}^{q}\PE\bigl[\|\sum\nolimits_{p=j}^{q} \nabla_{z_j}\partial_{z_{1}}^{\bigK_{I}}[f\circ G_{p}](x,\chunk{Z}{1}{p})\|^{2}\bigr].
\end{equation}
where for a non-empty subset $I\subseteq \{1,\ldots,d\}$, we denote $\bigK_{I}=K(\indiacc{1 \in I}\,\ldots, \indiacc{d \in I})$.
\end{proposition}
\begin{proof}
The proof is postponed to \Cref{proof:prop:main}.
\end{proof}

We aim at applying our main result to the estimation of expectations under the stationary distribution of ergodic diffusion processes. Let $b(x) = (b_1(x),\dots,b_d(x))$ be a drift function, $(\W_t)_{t \geq 0}$ be a $d-$dimensional Wiener process and assume that the stochastic differential equation
\begin{equation}
\label{eq:continuous_diffusion}
\rmd \X^x_t = -b(\X^x_t)\,\rmd t + \rmd \W_t, \, \X_0 = x.
\end{equation}
admits a unique strong solution $(\X^{x}_t)_{t \geq 0}$ for any $x \in \rset^d$. We consider the Euler-Maruyama discretization of the SDE \eqref{eq:continuous_diffusion}, i.e. the homogeneous Markov chain $(X_k^{x})_{k \geq 0}$, starting from $X_0^{x} = x \in \rset^d$
and defined by the following recursion: for any $k \in \nset$,
\begin{equation}
\label{eq:euler_discretized_diffusion}
X^x_{k+1} = X^x_{k} - \gamma b(X^x_k) + \sqrt{\gamma}Z_{k+1},
\end{equation}
where $\gamma > 0$ is a stepsize and $(Z_k)_{k \in \nset}$ is a sequence of i.i.d. $d-$dimensional Gaussian random variables with zero mean and identity covariance matrix. Note that the recurrence \eqref{eq:euler_discretized_diffusion} is a particular case of the general scheme \eqref{eq:chain_norm} with $\Phi(x,z) = x - \gamma b(x) + \sqrt{\gamma}z$. We impose some standard technical conditions on the drift function $b$, following \cite{debortoli2020}, namely,
\begin{assumption}
\label{assum:Lip_drift}
There exist a constant $L > 0$, such that $\|b(x) - b(y)\| \leq L \|x-y\|$ for any $x,y \in \rset^d$.
\end{assumption}
\begin{assumption}
\label{assum:Convexity_drift}
There exist a constant $m > 0$, such that $\ps{b(x) - b(y)}{x-y} \geq m \|x-y\|^2$ for any $x,y \in \rset^d$.
\end{assumption}
Under the assumptions \Cref{assum:Lip_drift} and \Cref{assum:Convexity_drift}, \Cref{prop:main} can be used to bound the variance of additive functionals of the Markov chains of the form \eqref{eq:euler_discretized_diffusion}.

\begin{theorem}
\label{prop:var_bounds}
Let $(X_{k}^{x})_{k \geq 0}$ be a Markov chain given by the recurrence \eqref{eq:euler_discretized_diffusion}, and assume that \Cref{assum:regularity_f}, \Cref{assum:Lip_drift}, and \Cref{assum:Convexity_drift} hold. Let $K \in \nset$. Assume in addition that there exist constants $C_f$ and $C_b$, such that for any $x \in \rset^d$, any multi-index $\kind \in \nset_0^{d}$ with $0 < \|\kind \| \leq K$, and any $u \in \{1,\dots,d\}$,
\[
|f^{(\kind)}(x)| \leq C_f, \quad |b_{u}^{(\kind)}(x)| \leq C_b\eqsp. %\quad \tcr{m_b \Id \preceq \Jac_{b}(x) \preceq M_b \Id}\eqsp,
\]
%where $\Jac_{b}(x)$ stands for Jacobian of $b$.
Then, for $0 < \gamma < \min(1/C_{b}, m/\ltwo)$ and any $n \in \nset$,
\[
\PVar\bigl[\pi_{n}^{(x,K)}(f)\bigr] \lesssim \frac{\gamma^{K-2}}{n}.
\]
Moreover, with the truncation point $\ntrunc(\gamma) = \lceil K \log{\gamma^{-1}}/(2m\gamma)\rceil$, variance of the truncated estimate \(\pi_{n,\ntrunc(\gamma)}^{(x,K)}(f)\) can be bounded as
\[
\PVar\bigl[\pi_{n,\ntrunc(\gamma)}^{(x,K)}(f)\bigr] \lesssim \frac{\gamma^{K-2}}{n},
\]
where \(\lesssim\) stands for inequality up to a constant not depending on $\gamma$ and $n$.
%depending on \(C_{f},\) \(C_{\Phi}\) and \(\kappa.\)
\end{theorem}
\begin{proof}
The proof is postponed to \Cref{proof:prop:var_bounds}.
\end{proof}
Theorem~\ref{prop:var_bounds} shows that under some conditions the variance of the estimate \(\pi^{(x,K)}_n(f)\) in the diffusion case \eqref{eq:euler_discretized_diffusion} satisfies
\[
\PVar\bigl[\pi_{n}^{(x,K)}(f)\bigr] \lesssim \frac{\gamma^{K-2}}{n}.
\]
At the same time, the variance of the standard Monte Carlo estimate \(\pi^{x}_n(f)\) is of order \(1/(n\gamma)\) and this order can not be reduced in general, see Example~\ref{ex:full_decomposition}.
Thus, for \(K\geq 2\) and \(\gamma\) small enough we have a clear variance reduction effect.

\begin{remark}
\label{eq:corollary_ULA} In the particular case of the Unadjusted Langevin algorithm (Example~\ref{ex:discr_diffusion}), assumptions of the Proposition~\ref{prop:var_bounds} can be verified for the smooth and strongly convex potential \(U\), that is, for $U \in C^2(\rset^d)$ and
\[
m_U \|x\|^2 \leq \ps{D^2U(y)x}{x} \leq M_U \|x\|^2
\]
for some $m_U > 0,\,M_U > 0$, and any $x, y \in \rset^d$.
\end{remark}

\section{Numerical experiments}
\label{sec:numerics}
In this chapter we evaluate our MAD-CV control variates on different model examples. Code to reproduce the experiments is available at \url{https://github.com/svsamsonov/MAD-CV}.
\subsection{Example~\ref{ex:full_decomposition} (continue)}
\begin{figure}[tbh]
\centering
\subfloat{\includegraphics[width=1.0\textwidth]{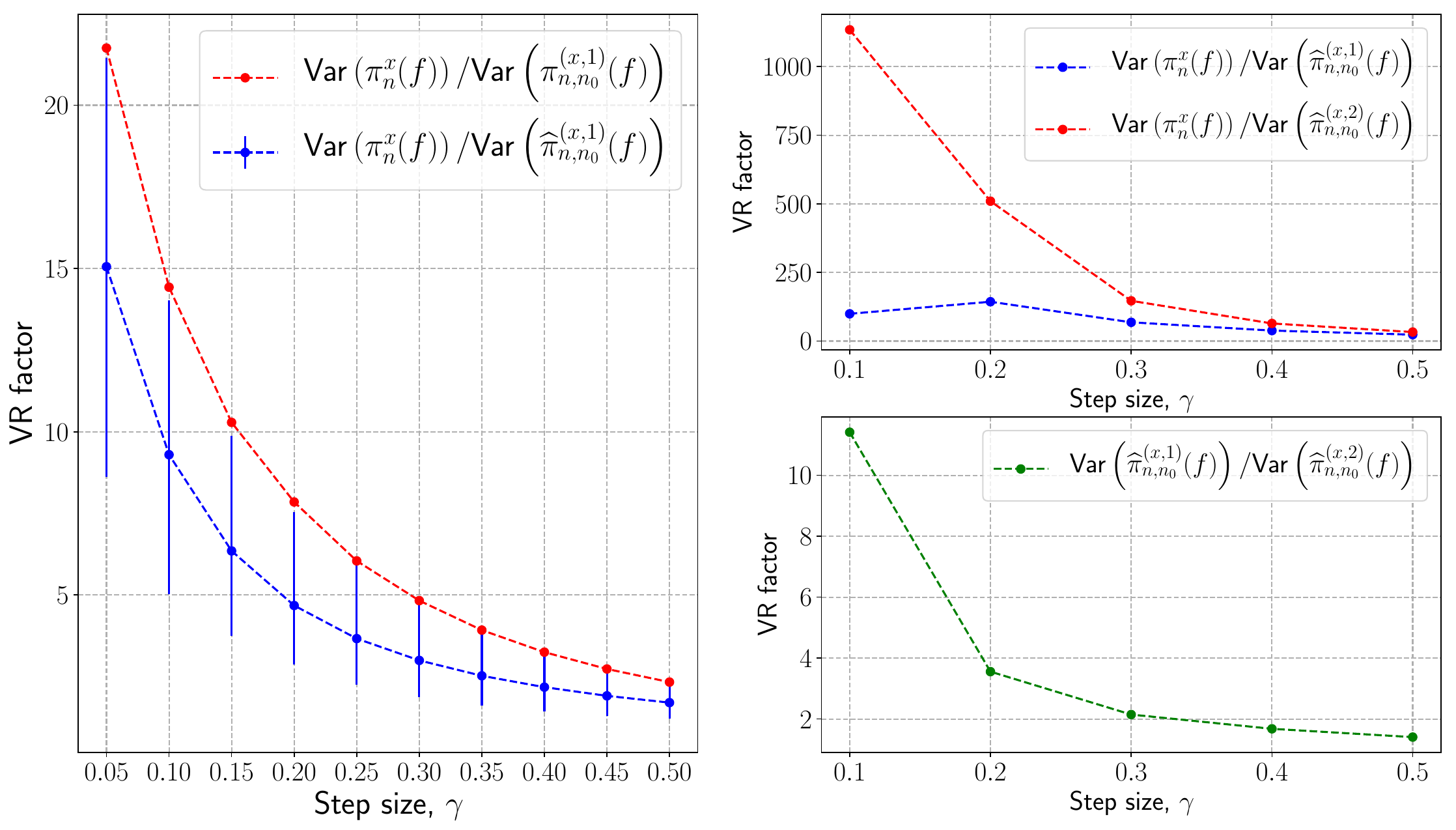}}
\caption{Left panel: variance reduction factors for the estimates $\pi_{n,\ntrunc}^{(x,1)}(f)$ (see \eqref{eq:VR-estimator}) and $\widehat{\pi}^{(x,1)}_{n,\ntrunc}(f)$ (see \eqref{eq:vr_estimator_truncated}) with respect to usual estimator $\pi_{n}^{x}(f)$ in \Cref{ex:full_decomposition}. We fix $f(y) = y^2$, $n = 10^4$, $X_0 = x = 1$ and consider different step size $\gamma \in [0.05,0.5]$. Right panel: same variance factors for the estimators $\hat{\pi}^{(x,1)}_{n,\ntrunc}(f)$, $\hat{\pi}^{(x,2)}_{n,\ntrunc}(f)$ with $f(y) = \sin{y}$ and variance ratio $\mathsf{Var}\bigl[\hat{\pi}^{(x,1)}_{n,\ntrunc}(f)\bigr]/\mathsf{Var}\bigl[\hat{\pi}^{(x,2)}_{n,\ntrunc}(f)\bigr]$.
\label{fig:gaus_ar_1_plot}}
\end{figure}

In this subsection we complete the Example~\ref{ex:full_decomposition} by evaluating the estimator $\hat{\pi}^{(x,1)}_{n,\ntrunc}(f) = \pi^x_n(f) -n^{-1}\hat{M}^x_{n,1,\ntrunc}$. Recall that we consider samples from the Gaussian distribution with  density $\pi(x) = (\sqrt{\pi})^{-1}\rme^{-x^2}$ using the ULA algorithm (see \eqref{eq:ULA_chain_gaussian}) and take $f(x) = x^2$.  We use different step sizes $\gamma \in (0.05,0.5)$ and sample training trajectory of length $5 \times 10^4$ for each step size. We solve the least squares  problems \eqref{eq:Qregr} with basis $\{1,x,x^2\}$ and the truncation points $\ntrunc = 5 + \lceil \log{\gamma^{-1}}/(4\gamma)\rceil$. Then we construct the control variate $\widehat M^{(x,1)}_{n,\ntrunc}$ defined in \eqref{eq:mad_cv_martingale_def}. Our goal is to compare the variance of the truncated estimator $\hat{\pi}^{(x,1)}_{n,\ntrunc}(f)$ to the variance of the "perfect" (with exact coefficients $\bar a$) estimator $\pi^{(x,1)}_{n,\ntrunc}(f)$. To this end, we we show two quantities. The first one is the ratio $\mathsf{Var}\left[\pi^{x}_{n}(f)\right] \Bigl/ \mathsf{Var}\left[\pi^{(x,1)}_{n,\ntrunc}(f)\right]$, which can be computed analytically for different $\gamma \in [0.05,0.5]$. The second one is the sample counterpart of the ratio $\mathsf{Var}\left[\pi^{x}_{n}(f)\right] \Bigl/ \mathsf{Var}\left[\hat{\pi}^{(x,1)}_{n,\ntrunc}(f)\right]$, computed over $100$ independent replications of $100$ test trajectories, each of length $n = 1 \times 10^{4}$. Left panel of \Cref{fig:gaus_ar_1_plot} contains error bars for $\mathsf{Var}\left[\pi^{x}_{n}(f)\right] \Bigl/ \mathsf{Var}\left[\hat{\pi}^{(x,1)}_{n,\ntrunc}(f)\right]$ and indicates that the use of regression   does not lead to a significant drop in the algorithm's performance.
\par
Next we aim to illustrate the results of \Cref{prop:var_bounds} by comparing $\mathsf{Var}\bigl[\hat{\pi}^{(x,1)}_{n,\ntrunc}(f)\bigr]$ to $\mathsf{Var}\bigl[\hat{\pi}^{(x,2)}_{n,\ntrunc}(f)\bigr]$. We fix $f(y) = \sin{y}$ and use different step sizes $\gamma \in [0.1,0.5]$. The least squares  problems \eqref{eq:Qregr} are solved with the regressors $\{1,x,x^2,x^3,x^4\}$. Following \Cref{prop:var_bounds}, we set the truncation points $\ntrunc = 10 + \lceil \log{\gamma^{-1}}/(2\gamma)\rceil$ for $\hat{\pi}^{(x,1)}_{n,\ntrunc}(f)$ and $\ntrunc = 10 + \lceil \log{\gamma^{-1}}/\gamma\rceil$ for $\hat{\pi}^{(x,2)}_{n,\ntrunc}(f)$, respectively. We compute the sample variance reduction factors $\mathsf{Var}\bigl[\pi^{x}_n(f)\bigr]\Bigl/ \mathsf{Var}\bigl[\hat{\pi}^{(x,K)}_{n,\ntrunc}(f)\bigr], K = 1,2$ over $100$ independent trajectories of length $2 \times 10^3$, repeat this procedure $100$ times and report the averaged variance reduction factors in the upper right panel of \Cref{fig:gaus_ar_1_plot}. To highlight the gain of the estimate $\hat{\pi}^{(x,2)}_{n,\ntrunc}(f)$, we report on the lower right panel of \Cref{fig:gaus_ar_1_plot} the averaged ratios $\mathsf{Var}\bigl[\hat{\pi}^{(x,1)}_{n,\ntrunc}(f)\bigr]/\mathsf{Var}\bigl[\hat{\pi}^{(x,2)}_{n,\ntrunc}(f)\bigr].$ Note that they scale approximately as $1/\gamma$, as predicted by \Cref{prop:var_bounds}.

\subsection{Comparison with vanilla ULA}
\begin{figure}[tbh]
\centering
\subfloat{\includegraphics[width=1.0\textwidth]{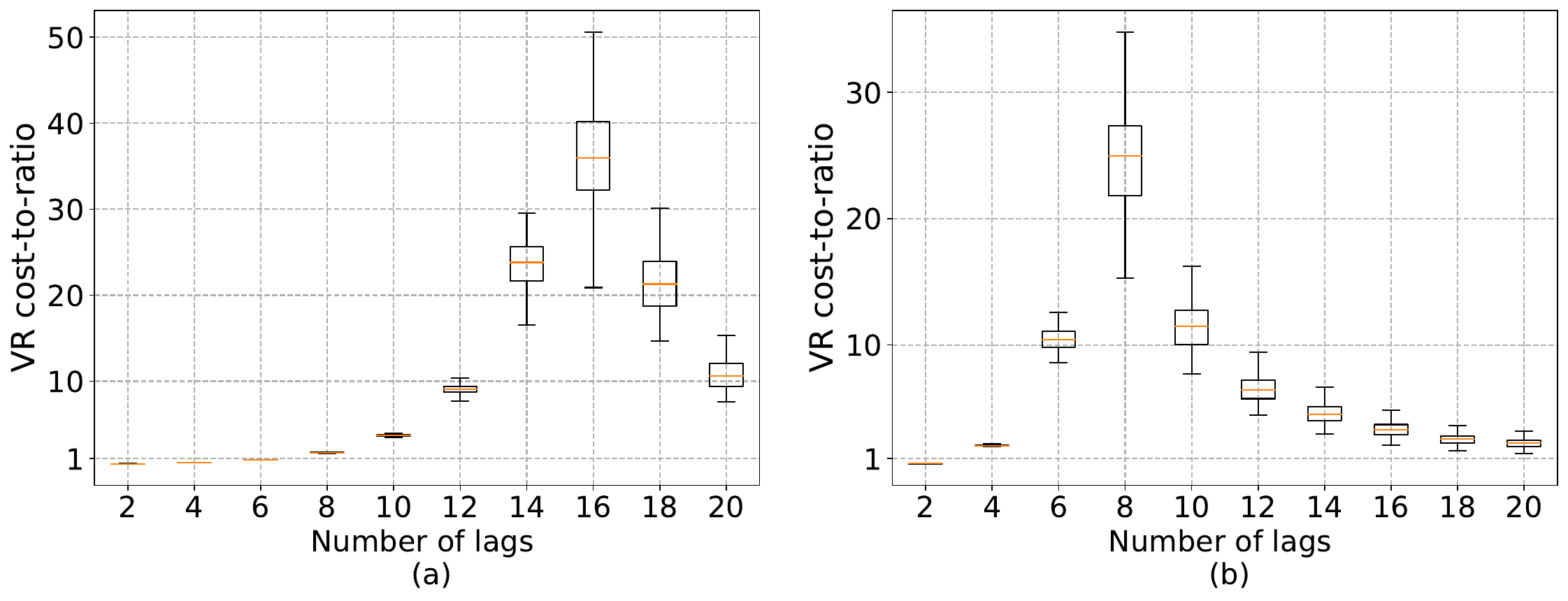}}
\caption{Cost-to-variance ratios \eqref{eq:cvR} as  functions of the truncation level $\ntrunc$ for  a two-dimensional standard Gaussian distribution. Subfigure $(a): f(x) = x_1 + x_2$, subfigure $(b): f(x) = x_1^2 + x_2^2$.\label{fig:gaus_cost}}
\end{figure}

\begin{figure}[tbh]
\centering
\subfloat{\includegraphics[width=1.0\textwidth]{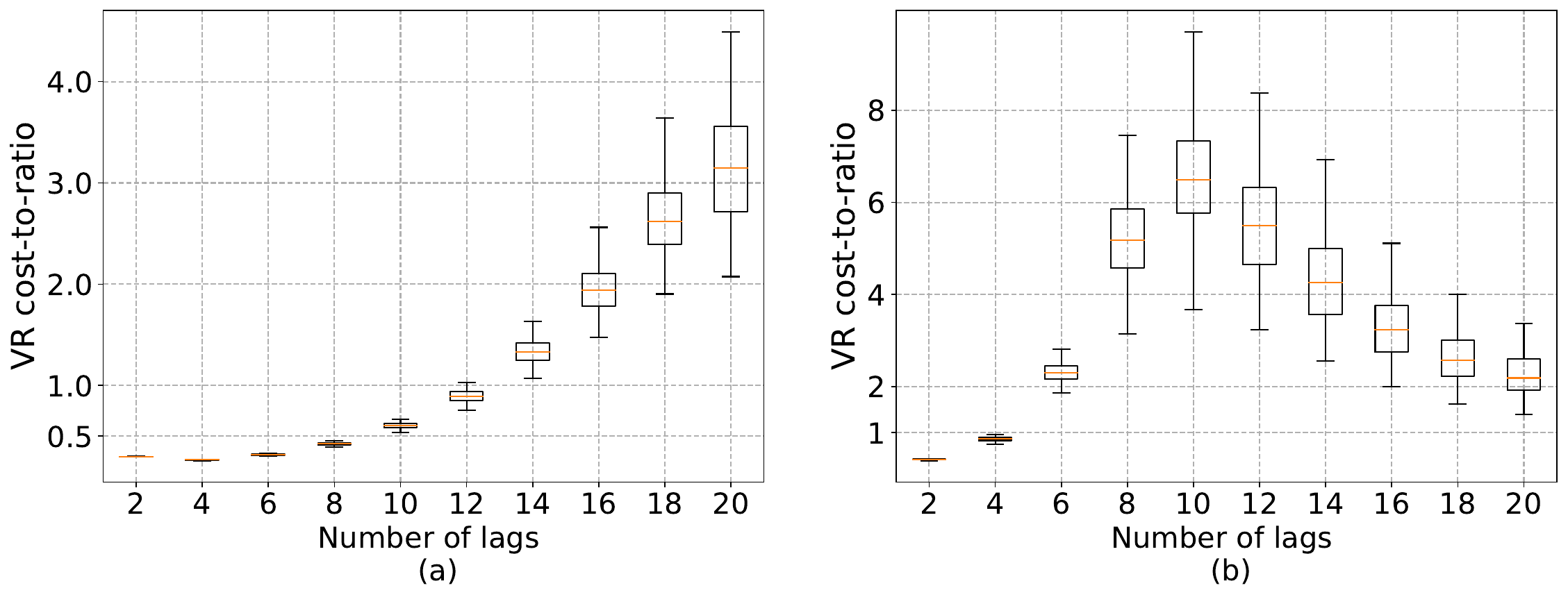}}
\caption{Cost-to-variance ratios  \eqref{eq:cvR} as  functions of the truncation level $\ntrunc$ for the mixture \eqref{eq:gmm} of two-dimensional Gaussian distributions. Subfigure $(a): f(x) = x_1 + x_2$, subfigure $(b): f(x) = x_1^2 + x_2^2$.\label{fig:gmm_cost}}
\end{figure}

We compare the variance reduction versus cost achieved by MAD-CV against plain Monte Carlo
for the ULA algorithm. We consider samples generated by ULA, where $\pi$ is either the standard normal distribution in dimension $d$ or the mixture of two $d-$dimensional standard Gaussian distributions of the form
 \begin{equation}
 \label{eq:gmm}
 \pi(x) = \frac{1}{2\sqrt{(2\pi)^{d}}} \left( \rme^{-(1/2)\|x-\mu\|^2} + \rme^{-(1/2)\|x+\mu\|^2}\right).
 \end{equation}
We fix $d = 2$ and $\mu = (0.5,0.5)$. In both examples, our goal is to estimate $\pi(f)$ with $f(x) = x_1+x_2$ %$f(x)=\sum_{i=1}^d x_i$
and %$f(x) = \sum_{i=1}^d x^2_i$
$f(x) = x_1^2 + x_2^2$.
We use a constant step size $\gamma=0.2$ and sample training trajectory of length $5 \times 10^4$ with the starting point $X_0 = x = (1,1)$. Then we solve the least squares  problems \eqref{eq:Qregr} with the class of regressors
%\[
%\Psi_2 = \beta_0 + \beta_1 x_1 + \beta_2 x_2 + \beta_3 x_1^2 + \beta_4 x_1x_2 + \beta_5 x_2^2,
%\]
$\{x_1,x_2,x_1^2,x_1x_2,x_2^2\}$
for the different choices of truncation point $\ntrunc \in [2,20]$.
We construct the control variate $M_{n,\ntrunc}^{(x,K)}$, defined in \eqref{eq:multidim_decomposition}.
%That is, for $K=1$ we compute only coefficients, corresponding to the Hermite polynomials
%\[
%\mathbf{H}_{(1,0)}(x) = H_{1}(x_1), \, \mathbf{H}_{(0,1)}(x) = H_{1}(x_2).
%\]
%For $K=2$ we additionally compute coefficients corresponding to the polynomials
%\[
%\mathbf{H}_{(2,0)}(x) = H_{2}(x_1), \, \mathbf{H}_{(0,2)}(x) = H_{2}(x_2).
%\]
We finally estimate the cost-to-variance ratio (degree of variance reduction relative to costs) as follows
\begin{eqnarray}
\label{eq:cvR}
\mathcal{R}(f,K,n,\ntrunc)=\frac{\mathrm{cost}\{ \pi^{x}_{n}(f)\}\PVar[\pi^{x}_{n}(f)]}{\mathrm{cost}\{ \pi^{(x,K)}_{n,\ntrunc}\} \PVar[\pi^{(x,K)}_{n,\ntrunc}(f)]}
\end{eqnarray}
by its empirical counterpart, computed over $100$ independent trajectories, each of length $n = 5 \times 10^4$. Note that for $2-$dimensional standard Gaussian vector $Z = (Z_1,Z_2)$, for multi-indices $\kind = (k_1,k_2) \in \{(2,1),(1,2),(2,2)\}$ it holds that
\[
\PE[\psi(x - \gamma U(x) + \sqrt{2\gamma}Z)H_{k_1}(Z_1)H_{k_2}(Z_2)] = 0,
\]
for any $\psi(x) \in \{x_1,x_2,x_1^2,x_1x_2,x_2^2\}$. This implies that the coefficients
$\bar{a}_{r,\kind}(x) = 0$, for any $x \in \rset^{d}$ and $\kind = (k_1,k_2) \in \{(2,1),(1,2),(2,2)\}$. Since for a fixed $K$ the cost of computing $\pi^{x}_{n}(f)$ is proportional to the cost of computing function $f$, we set for $K = 1$
 \begin{equation*}
 \mathrm{cost}\{ \pi^{(x,1)}_{n, \ntrunc}(f)\} = \mathrm{cost}\{ \pi^{x}_{n}(f)\} \times \ntrunc \times 3,
 \end{equation*}
since for the fixed $r$, each coefficient $\bar{a}_{r,\kind}(y)$  is a polynomial, which can be computed at the same cost as $f$. Similarly, for $K = 2$ we set
  \begin{equation*}
 \mathrm{cost}\{ \pi^{(x,2)}_{n, \ntrunc}(f)\} = \mathrm{cost}\{ \pi^{x}_{n}(f)\} \times \ntrunc \times 5,
 \end{equation*}
since we need to evaluate $5\ntrunc$ coefficients in addition to each evaluation of $f$. Variance reduction costs for Gaussian distribution and different truncation points $\ntrunc$ are summarized in Figure~\ref{fig:gaus_cost}, and for the Gaussian mixture - in Figure~\ref{fig:gmm_cost}. Note that for both examples MAD-CV allows us to obtain a significant gain in terms of cost-to-variance ratios.

\subsection{Random Walk Metropolis (RWM) example}
We illustrate the application of MAD-CV to the RWM algorithm.
RWM is an MCMC algorithm using random walk proposal.
Let $\{ U_p \}_{p=1}^\infty$ and $\{ Z_p \}_{p=1}^\infty$ be independent  i.i.d. sequences,
with $U_{p} \sim \mathrm{Unif}[0,1]$ and $Z_{p} \sim \mathcal{N}(0,\Id_{d})$. Then the $p$-th RWM iterate writes as
\begin{equation}
\label{eq:MH_alg}
X_{p+1}^{x} = X_{p}^{x} + \mathbbm{1}\bigl\{U_{p+1}\leq \alpha(X_{p}^{x},X_{p}^{x}+\sqrt{\gamma}Z_{p+1})\bigr\}(\sqrt{\gamma}Z_{p+1})\,,
\end{equation}
where $\alpha(x,y) = \min\bigl\{1, \pi(y)/\pi(x)\bigr\}$ is the acceptance ratio. In this experiment $\pi$ is set to be the standard normal distribution in dimension $d = 2$.
We aim to estimate $\pi(f)$ with $f(x) = x_1^2 + x_2^2$, using RWM. The variance of the incremental distribution is determined by $\gamma = 1.0$, which leads to an acceptance rate in stationarity of approximately~$0.55$. We sample a training trajectory of length $N = 10^{6}$, and solve the regression problem \eqref{eq:Qregr} with the polynomial regressors $x_1^{d_1}x_{2}^{d_2}$, $d_1 + d_2 \leq 4$. We illustrate that
MAD-CV can benefit when taking into account both randomness in $U_p$ and $Z_p$. Namely, for a multi-index $\kind = (k_1,k_2,k_3)$, $z \in \rset^2$ and $u \in \rset$, we consider basis functions
\begin{equation}
\label{eq:phi_legendre_hermite}
\psi_{\kind}(z,u) = H_{k_1}(z_1)H_{k_2}(z_2)P_{k_3}(u)
\end{equation} with $(P_k)_{k \in \nset}$ being shifted Legendre polynomials on $[0,1]$. We use QMC to evaluate the corresponding functions $\widehat a_{r+1,\kind}$ in \eqref{eq:a-pol-gen}. We write $\widehat{\pi}_{n,n_0}^{(x,\mathbf{K})}(f), \mathbf{K} = (K_1,K_2,K_3)$ for the version of estimator \eqref{eq:vr_estimator_truncated} based on the coefficients $\widehat a_{r+1,\kind}$, $\kind \leq \mathbf{K}$.

\begin{figure}[tbh]
\centering
\subfloat{
\includegraphics[width=0.47\textwidth]{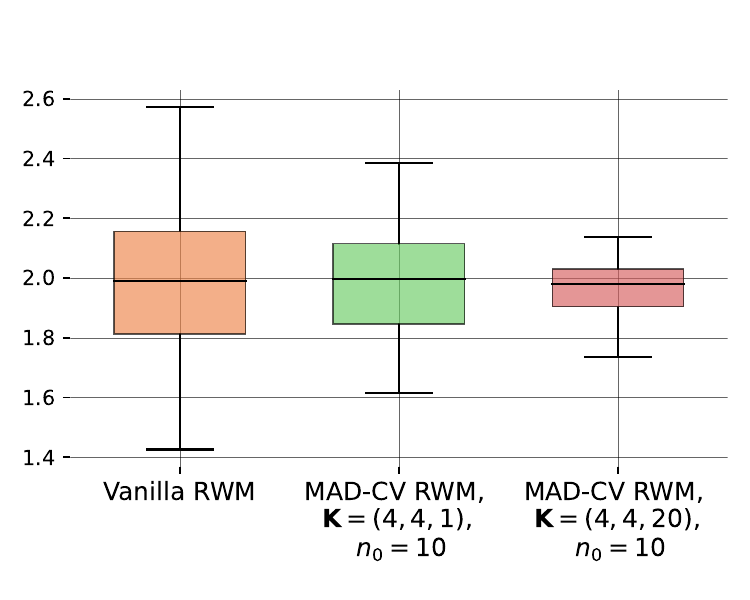}
\label{fig:subfig1}}
\qquad
\subfloat{
\includegraphics[width=0.45\textwidth]{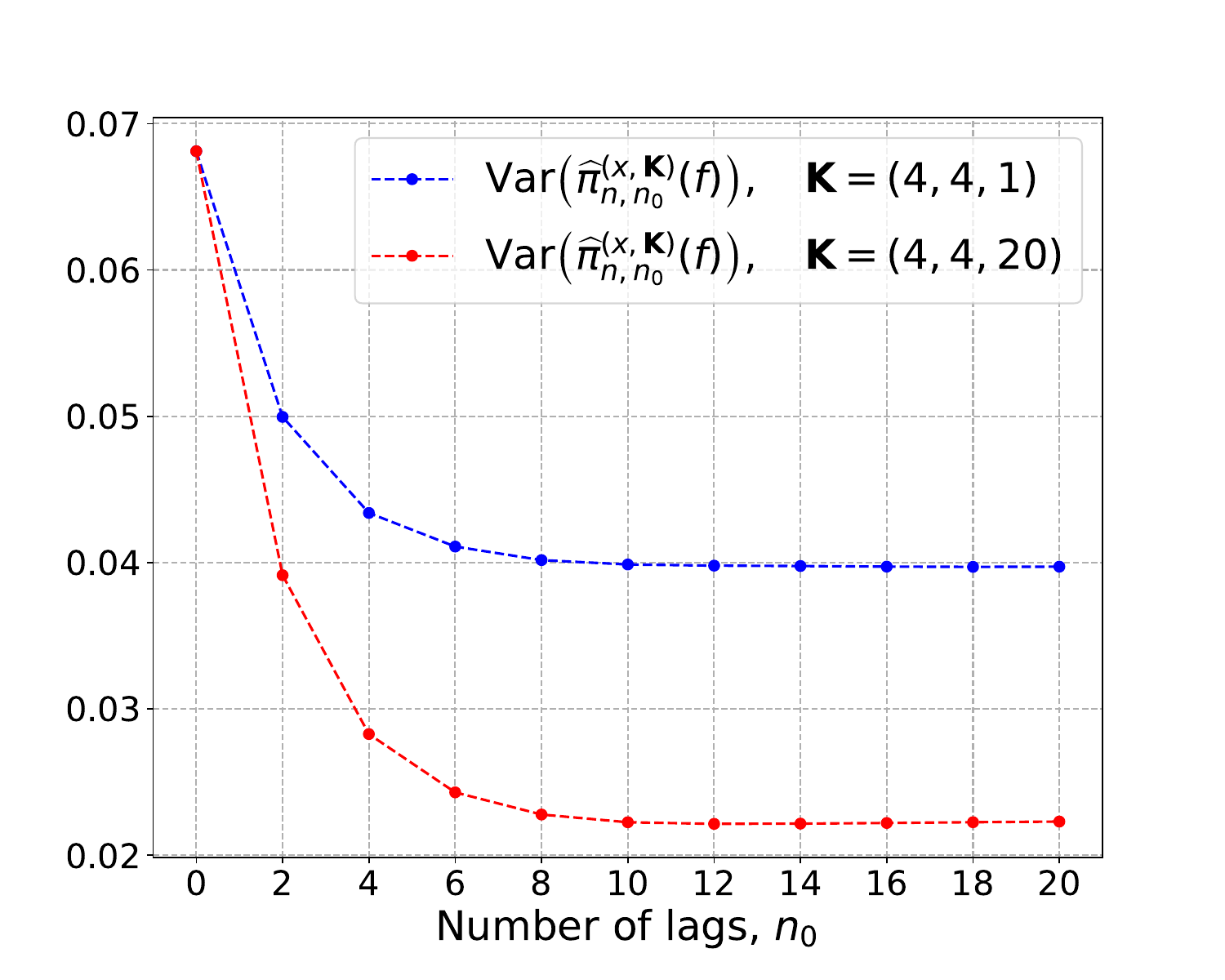}
\label{fig:subfig2}}
\caption{Left panel: boxplots of ergodic averages for RWM algorithm in the case of $2$-dimensional Gaussian distribution. Right panel: respective sample variances of the estimators $\widehat{\pi}_{n,n_0}^{(x,\mathbf{K})}(f)$ for different truncation points $n_0$ (computed over $100$ independent runs).
\label{fig:RWM_gaussian}}
\end{figure}

To test our variance reduction algorithm, we generate \(100\) independent trajectories of length $n = 1 \times 10^3$. We use $4$ Hermite polynomials in each coordinate (that is, $k_1,k_2 \in \{1,\dots,4\}$). The compared cases are when the MAD-CV are only applied to the proposal \eqref{eq:MH_alg}  (that is, $\mathbf{K} = (4,4,1)$), and when the MAD-CV are applied jointly on the proposal and the acceptance step ($\mathbf{K} = (4,4,20)$). \Cref{fig:RWM_gaussian} displays the boxplots of the estimates together with the estimated standard deviations of the corresponding estimates $\widehat{\pi}^{(x,\kind)}_{n,n_0}$ for different truncations $n_0 \in \{2,\dots,20\}$. Note that combining Legendre and Hermite polynomials allows to achieve better variance reduction compared to the case when only Hermite polynomials are used.

\subsection{Euler scheme for discretized diffusion}
\begin{figure}[tbh]
\centering
\subfloat[Subfigure 1 list of figures text][$f(x) = \sum\nolimits_{i=1}^{d}x_i$]{
\includegraphics[width=0.46\textwidth]{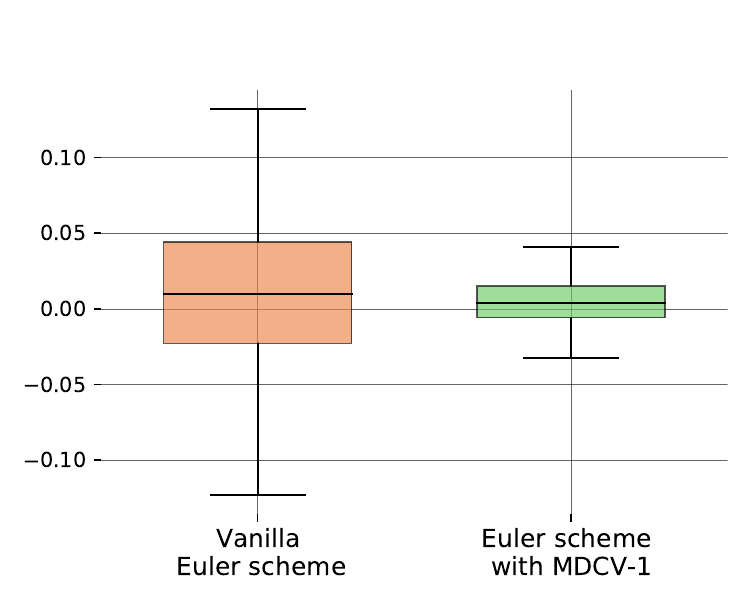}
\label{fig:subfig1}}
\qquad
\subfloat[Subfigure 2 list of figures text][$f(x) = \sum\nolimits_{i=1}^{d}x_i^2$]{
\includegraphics[width=0.46\textwidth]{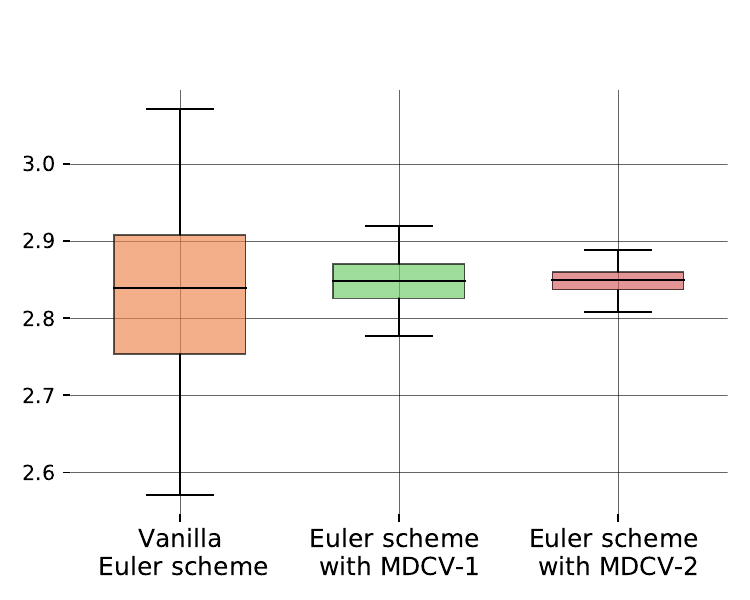}
\label{fig:subfig2}}
\caption{Boxplots of ergodic averages obtained via Euler scheme for the SDE \eqref{eq:diffusion_example}, $d = 5$. The compared estimators are the ordinary empirical average (Vanilla) and our method of martingale control variates (MAD-CV-$1$ and MAD-CV-$2$).
\label{fig:discretized_diffusion}}
\end{figure}

We consider the $d-$dimensional stochastic differential equation
\begin{equation}
\label{eq:diffusion_example}
d\X^x_t = -b(\X^x_t)\,\rmd t + \rmd \W_t, \, \X_0 = x \in \rset^d
\end{equation}
with the drift function
%\[
%b(x) = \bigl(x_1 + 0.5 \sin{x_2}, x_2 + 0.5 \sin{x_1}\bigr)^{\top}, \, x = (x_1,x_2)^{\top} \in \rset^2.
%\]
\[
b(x) = (b_i(x))_{i=1}^{d}, \quad b_i(x) = x_i + a \sin{x_{1 + i \bmod d}}, \quad x = (x_1,\dots,x_d)^{\top} \in \rset^d, \quad a \in \rset
\]
We aim at estimating $\pi(f)$ for the functions $f(x) = \sum\limits_{i=1}^{d}x_i$ and $f(x) = \sum\limits_{i=1}^{d}x_i^2$, where $\pi$ is an ergodic distribution of \eqref{eq:diffusion_example}. We fix $d = 5$, $a = 0.5$, and consider the Euler-Maruyama discretization of \eqref{eq:diffusion_example} with constant stepsize $\gamma = 0.1$, and approximate $\pi(f)$ by $\pi^{x}_n(f)$ and its MAD-CV counterparts. Note that the assumptions of \Cref{prop:var_bounds} are satisfied. We consider estimators $\pi_{n,n_0}^{(K)}(f)$ with $n_0 = 20$ and $K = 1$ or $K = 2$. We refer to them as to MAD-CV-1 and MAD-CV-2, respectively. First we sample a training trajectory of length $N = 10^{4}$, and solve the regression problem \eqref{eq:Qregr} with the class of regressors $\{x_i, x_jx_k\}$ for $i,j,k \in \{1,\dots,d\}$.
%\[
%\Psi_2 = \beta_0 + \beta_1 x_1 + \beta_2 x_2 + \beta_3 x_1^2 + \beta_4 x_1x_2 + \beta_5 x_2^2.
%\]
To test our variance reduction algorithm, we generate $100$ independent test trajectories of length $5 \times 10^3$ and compute $\pi^{x}_n(f)$ and $\pi^{(x,K)}_{n,n_0}(f)$. The corresponding boxplots are presented in \Cref{fig:discretized_diffusion}. 

%We also illustrate the statement of \Cref{prop:var_bounds}. We fix $d=2$, $a = 0.2$, $\ntrunc = 50$, and consider $f(x) = \sin{x_1} + \sin{x_2}$. We are interested in the variance of $\pi^{(x,2)}_{n,n_0}(f)$ for different values of step size $\gamma$, used during the data generation process. We show the corresponding boxplots in the \Cref{fig:gamma_influence}. Note that they reflect much less dependence on the step size $\gamma$, compared to the vanilla Euler scheme estimates.

\subsection{Lotka-Volterra model with feedback control}
We consider the stochastic Lotka-Volterra predator-prey model with feedback control, following \cite{lotka_volterra_control}:
\begin{equation}
\label{eq:lv_discrete}
\begin{cases}
\rmd \X_1(t) = \X_1(t)\bigl(r_1 - a_{1,1}\X_1(t) - a_{1,2}\X_2(t)\bigr)\,\rmd t + \sigma_1 \X_1(t) \rmd \W_{1,t}, \quad \X_1(0) = x_{1,0} \in \rset_{+} \\
\rmd \X_2(t) = \X_2(t)\bigl(r_2 + a_{2,1}\X_1(t) - a_{2,2}\X_2(t) - c \X_3(t)\bigr)\,\rmd t + \sigma_2 \X_2(t) \rmd \W_{2,t}, \quad \X_2(0) = x_{2,0} \in \rset_{+} \\
\rmd \X_3(t) = \bigl(-e \X_3(t) + h \X_2(t)\bigr) \rmd t\eqsp, \quad \X_3(0) = x_{3,0} \in \rset_{+}
\end{cases}
\end{equation}
where $\W_{i,t} \, (i = 1,2)$ denote independent Wiener processes, parameters $a_{i,i} > 0$ correspond to intraspecific competition rates, $a_{i,j}, i \neq j$ stand for capturing rates of the prey and predator, $r_i > 0$ represent the intrinsic growth rate of the population and $\sigma_i^2 > 0$. We consider Euler-Maruyama discretisation of the equation \eqref{eq:lv_discrete} with step size $\gamma = 0.1$, and fix the hyperparameters
\[
A = \begin{pmatrix}
    0.2 & 0.2 \\
    -0.2 & 0.4 \\
\end{pmatrix}, \, r = (r_1,r_2) = (1.2,1.5), \, e = 0.5, \, h = 0.2, \, X(0) = x = (7,6,5)\eqsp.
\]
Note that the assumptions $(A_1)$ and $(A_2)$ from \cite{lotka_volterra_control}, namely $a_{1,1}a_{2,2} > a_{2,1}a_{1,2}$ and $r_1/r_2 > a_{1,1}/(a_{2,2} + ch/e)$ are satisfied, and the system \eqref{eq:lv_discrete} oscillates around its equillibrium point. We fix $f(x) = x_1$ or $f(x) = x_2$, and aim at approximating $\pi(f)$ by $\pi^{x}_n(f)$ and its variance-reduced counterparts. We sample a training trajectory of length $N = 5 \times 10^3$, and solve the regression problem \eqref{eq:Qregr} with the class of regressors $\{x_i, x_jx_k\}$ for $i,j,k \in \{1,\dots,d\}$. We set the truncation level $\ntrunc = 50$ and generate $100$ independent test trajectories of length $n = 5 \times 10^{3}$. For each trajectory we compute $\pi^{x}_n(f)$ and its variance-reduced counterpart $\pi_{n,\ntrunc}^{(x,1)}(f)$. We show the simulated trajectories of the system \eqref{eq:lv_discrete} in \Cref{fig:lv_subfig_1}, and the boxplots corresponding to $f(x) = x_1$ in  \Cref{fig:lv_subfig_2}.

\begin{figure}[tbh]
\centering
\subfloat[Subfigure 1 list of figures text][Simulated trajectories of \eqref{eq:lv_discrete}]{
\includegraphics[width=0.46\textwidth]{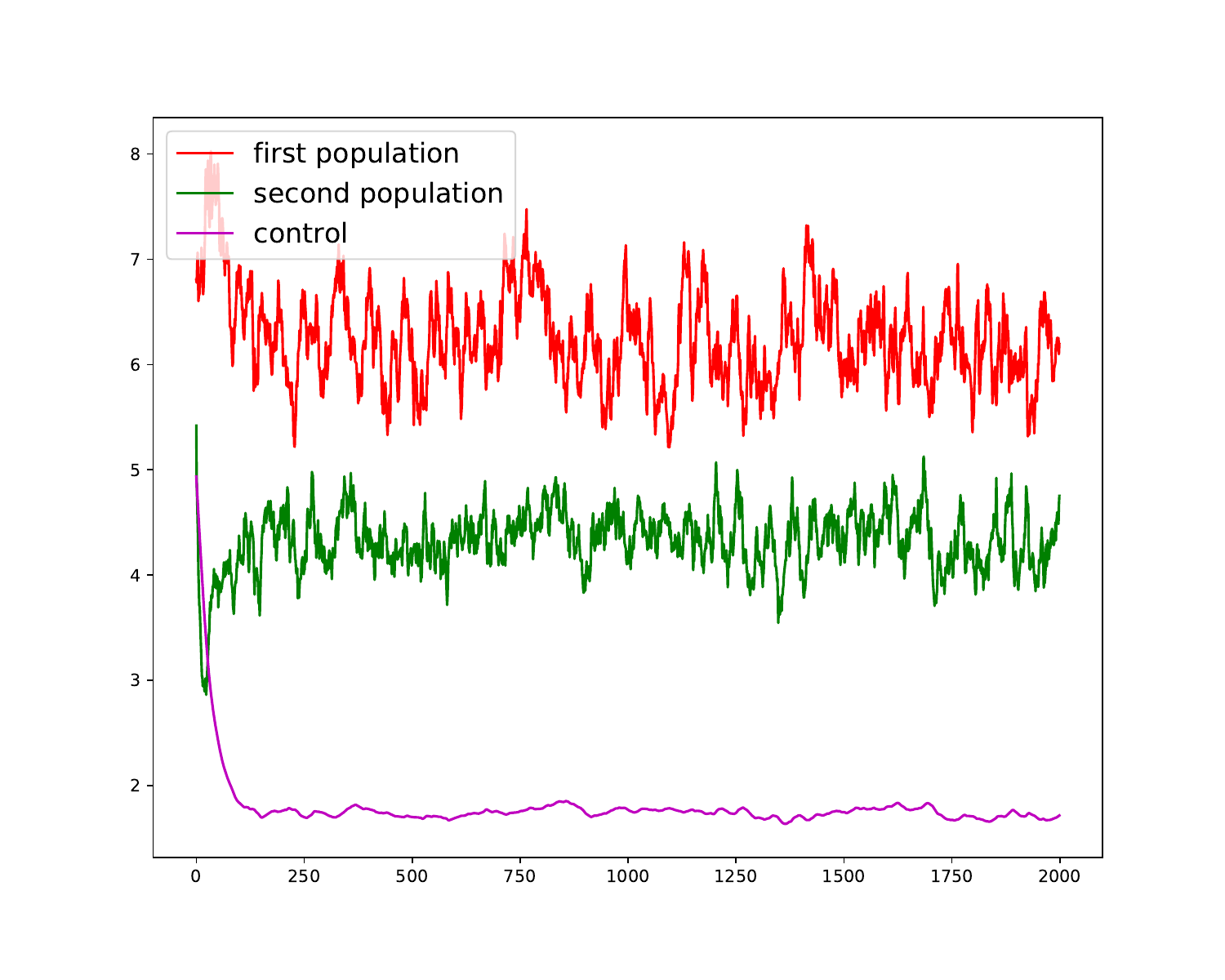}
\label{fig:lv_subfig_1}}
\qquad
\subfloat[Subfigure 2 list of figures text][Estimating $\pi(f)$ for $f(x) = x_1$]{
\includegraphics[width=0.46\textwidth]{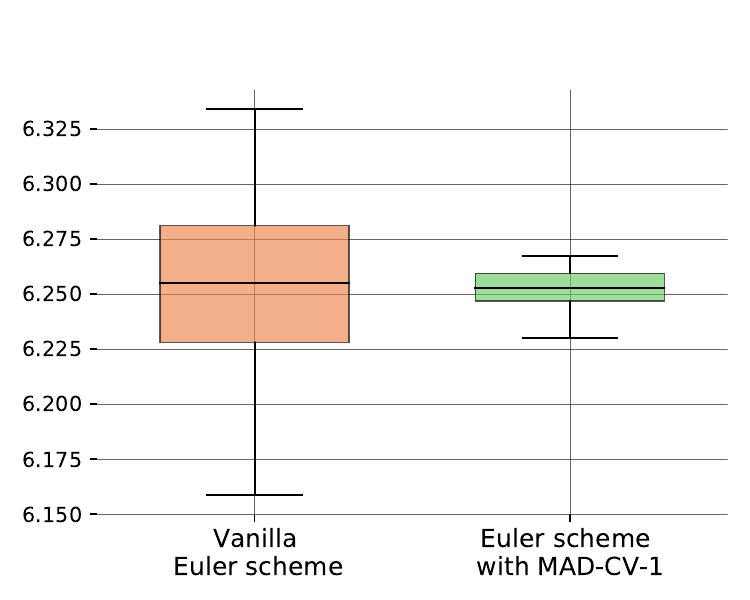}
\label{fig:lv_subfig_2}}
\caption{Simulated trajectories and estimates for Lotka-Volterra model with feedback control
\label{fig:LV_model_with_control}}
\end{figure}

\subsection{Multidimensional stochastic Lotka-Volterra model}
Following \cite{sabanis_lv}, we consider Lotka–Volterra model for a system with $d$ interacting components, corresponding to the case of facultative mutualism, namely
\begin{equation*}
\rmd x(t) = \diag\bigl(x_1(t),\dots,x_d(t)\bigr)\bigl[r + Ax(t)\bigr], \quad t \geq 0 \eqsp,
\end{equation*}
where
\[
x(t) = \bigl(x_1(t),\dots,x_d(t)\bigr)^{T}, \, r = (r_i) \in \rset^{d}, \, A = (a_{i,j}) \in \rset^{d \times d}
\]
Stochastically perturbing parameters $a_{i,j}$, we come up with the system
\begin{equation}
\label{eq:multidim_lv_system}
\rmd \X_{i}(t) = \X_i(t)\bigl(r_i + \sum\limits_{j=1}^{d}a_{i,j}\X_{j}(t)\bigr)\,\rmd t + \X_i(t)\bigl(\sum\limits_{j=1}^{d}\sigma_{i,j}\X_j(t)\bigr)\rmd \W_{i,t}, \quad i = 1,\dots,d
\end{equation}
where $(\W_{i,t})_{t \geq 0} \, (i = 1,\dots,d)$ denote independent Wiener processes and $\Sigma = (\sigma_{i,j}) \in \rset^{d \times d}$ is a matrix with $\sigma_{i,i} > 0$ and $\sigma_{i,j} \geq 0, i \neq j$. We consider the Euler-Maruyama discretisation of the equation \eqref{eq:multidim_lv_system} with constant step size $\gamma = 0.02$, and fix the hyperparameters
\[
A = \begin{pmatrix}
    -0.2 & -0.1 & -0.3 & -0.2 \\
    -0.1 & -0.2 & -0.1 & -0.3 \\
    -0.2 & -0.1 & 0.4 & 0 \\
    -0.1 & -0.3 & 0 & 0.4 \\
\end{pmatrix}, \, r = (r_i) = (4.0,3.0,1.0,0.5)^T, \, \X_{0} = x = (7,6,5,5)^T\eqsp.
\]
We also set $\sigma_{i,i} = 0.1, \, \sigma_{i,j} = 0, \, i \neq j$. Note that the conditions of \cite[Theorem~1]{sabanis_lv} are satisfied, and the system \eqref{eq:lv_discrete} has a unique positive solution. We fix $f(x) = x_1$, and aim at approximating $\pi(f)$ by $\pi^{x}_n(f)$ and its variance-reduced counterparts. We sample a training trajectory of length $N = 1 \times 10^4$, and solve the regression problem \eqref{eq:Qregr} with the class of regressors $\{x_i, x_jx_k\}$ for $i,j,k \in \{1,\dots,d\}$. We set the truncation level $\ntrunc = 50$ and generate $100$ independent test trajectories of length $n = 5 \times 10^{3}$. For each trajectory we compute $\pi^{x}_n(f)$ and its variance-reduced counterpart $\pi_{n,\ntrunc}^{(x,1)}(f)$. We show the simulated trajectories of the system \eqref{eq:lv_discrete} in \Cref{fig:lv_subfig_1}, and the boxplots corresponding to $f(x) = x_1$ in  \Cref{fig:lv_subfig_2}.

\begin{figure}[tbh]
\centering
\subfloat[Subfigure 1 list of figures text][Simulated trajectories of \eqref{eq:multidim_lv_system}]{
\includegraphics[width=0.46\textwidth]{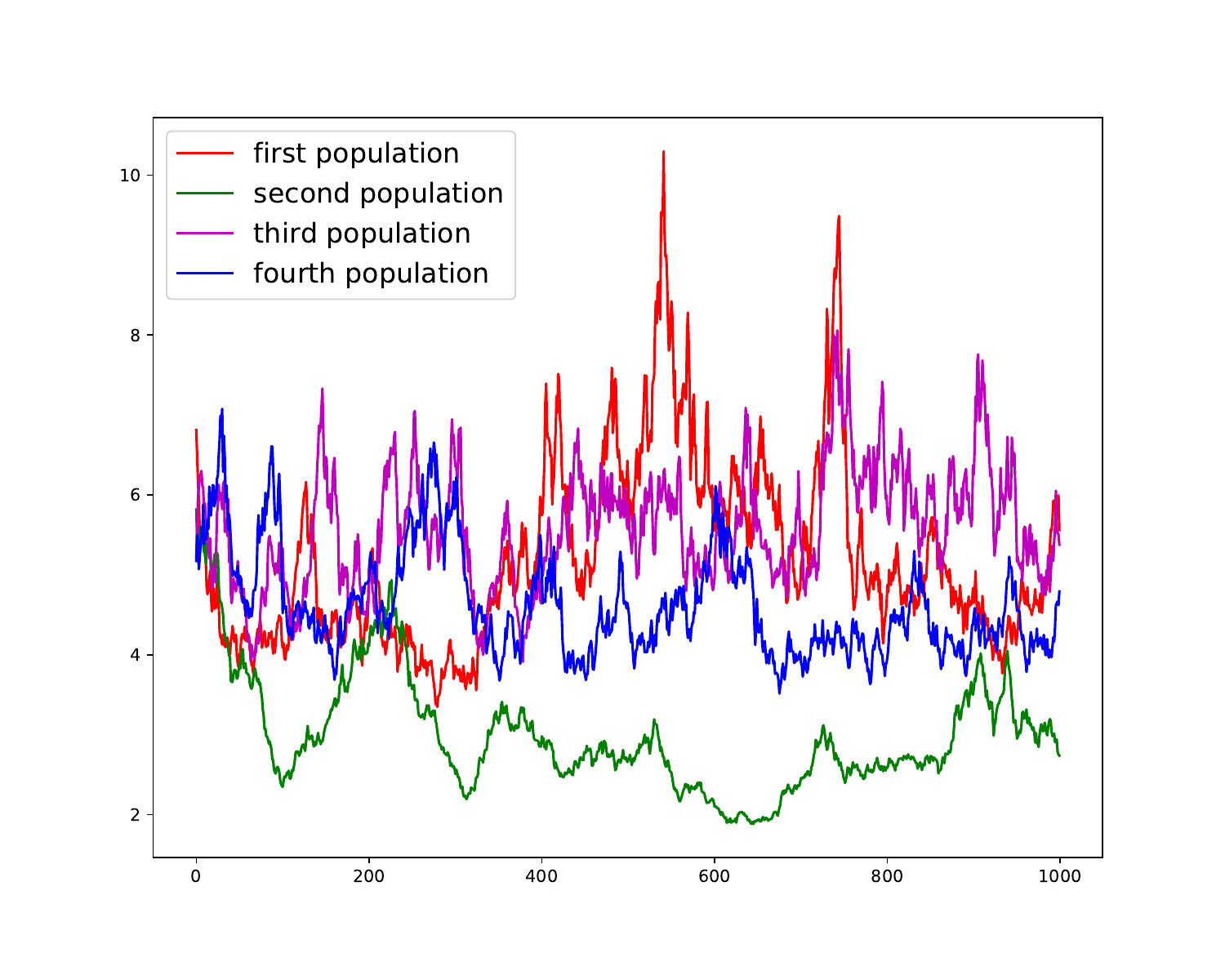}
\label{fig:lv_subfig_1}}
\qquad
\subfloat[Subfigure 2 list of figures text][Estimating $\pi(f)$ for $f(x) = x_1$]{
\includegraphics[width=0.46\textwidth]{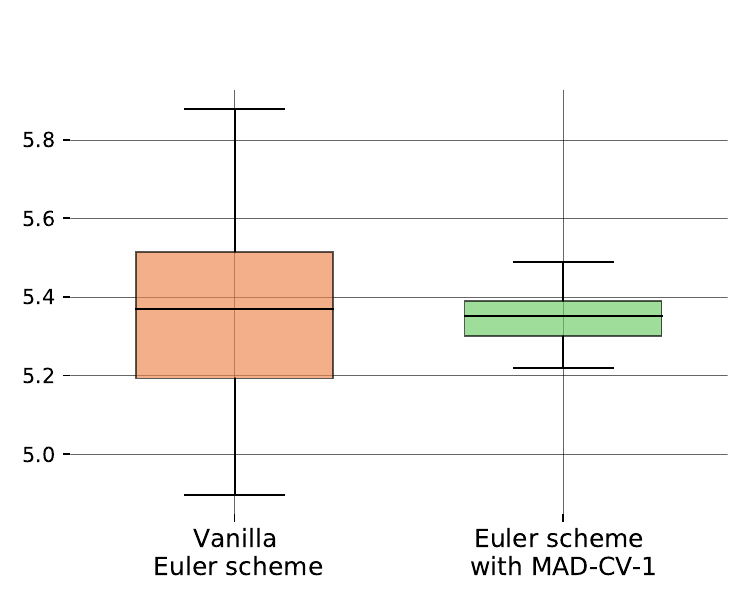}
\label{fig:lv_subfig_2}}
\caption{Simulated trajectories and estimates for multidimensional stochastic Lotka-Volterra model
\label{fig:LV_model_multidimensional}}
\end{figure}

\section{Proofs}
\label{sec:proofs}
\subsection{Notations.}\label{par:notations}
\begin{comment}
We use the notations $\nset=\{1,2,\ldots\}$ and $\nset_0=\mathbb N\cup\{0\}$. For the twice differentiable function $g: \rset^d \rightarrow \rset$ we denote by $D^2g(x)$ its Hessian at point $x$. For $m\in\mathbb N$, a smooth function
$h\colon\mathbb R^{d\times m}\to\mathbb R$
with arguments being denoted
$(y_1,\ldots,y_m)$, $y_i\in\mathbb R^d$, $i=1,\ldots,m$,
a multi-index $\mathbf k=(k_i)\in\mathbb N_0^d$,
and $j\in\{1,\ldots,m\}$,
we use the notation $\partial^{\mathbf k}_{y_j} h$ for the multiple derivative of $h$
with respect to the components of~$y_j$:
\[
\partial^{\mathbf k}_{y_j} h(y_1,\ldots,y_m)
:=\partial^{k_d}_{y_j^d}
\ldots
\partial^{k_1}_{y_j^1}
h(y_1,\ldots,y_m),
\quad y_j=(y_j^1,\ldots,y_j^d),
\]
where \(\partial^{k_s}_{y_j^s} h(y_1,\ldots,y_m) \) stands for partial derivative of order \(k_s\) for the function \(h\) with respect to the \(s\)th - coordinate of the vector \(y_j\).
\end{comment}
For multi-indices $\kind = (k_1,\dots,k_d)$ and $\lind = (l_1,\dots,l_d) \in \nset_{0}^{d}$, such that $|\kind| > 0$, $|\lind| > 0$, for $m \in \nset$ and smooth function $f: \rset^{d \times m} \rightarrow \rset^d$, such that
\[
f(\chunk{z}{1}{m}) := \bigl(f_1(\chunk{z}{1}{m}),\dots, f_d(\chunk{z}{1}{m})\bigr),\, z_i \in \rset^d, \, i \in \{1,\dots,m\}
\]
we write
\begin{equation}
\label{eq:vector_multiderivative}
\bigl[\partial^{\lind}_{z_i} f \bigr]^{\kind} := \prod\limits_{j=1}^{d}\bigl(\partial^{\lind}_{z_i} f_j
\bigr)^{k_j}\eqsp.
\end{equation}
We write $\Jac_{f}(\chunk{z}{1}{m}) \in \rset^{dm \times d}$ for the Jacobian of $f$ at point $\chunk{z}{1}{m}$, and $\nabla_{z_{i}}f \in \rset^{d \times d}$ for the matrix of partial derivatives $(\Jac_{f}^{(z_i)})_{u,v} = \partial_{z_{i,u}}f_v(\chunk{z}{1}{m})$. For the multi-indices $\kind, \lind \in \nset_{0}^{d}$, we write that $\kind \prec \lind$, if one of the following holds:
\begin{itemize}
    \item $|\kind| < |\lind|$;
    \item $|\kind| = |\lind|$ and $\kind_1 < \kind_1$, or
    \item $|\kind| = |\lind|$, $\kind_1 = \lind_1$, \dots, $\kind_m = \lind_m$, and $\kind_{m+1} = \lind_{m+1}$.
\end{itemize}
For the multi-indices $\qind, \rind \in \nset_{0}^{d}$ we define $P(\qind,\rind)$ the set of multi-indices $\kind_{i}, \lind_i \in \nset_{0}^{d}$,
%\[
%P(\qind,\rind) = \{(\kind_{1},\dots,\kind_{|\qind|},\lind_{1},\dots,\lind_{|\qind|})\},
%\]
such that for some $1 \leq s \leq |\qind|$, $\kind_i = 0$ and $\lind_i = 0$ for $1 \leq i \leq |\qind| - s$, $|\kind_i| > 0$ for $|\qind|-s+1 \leq i \leq |\qind|$ and $0 \prec \ell_{|\qind|-s+1} \prec \dots \prec \ell_{|\qind|}$ are such that
\begin{equation}
\label{eq:constraints}
\sum\limits_{i=1}^{|\qind|}\kind_i = \rind, \quad \sum\limits_{i=1}^{|\qind|}|\kind_i|\lind_i = \qind.
\end{equation}
\subsection{Proof of Theorem~\ref{thm:main-repr}}
\label{sec:proof:thm:main-repr}
The expansion obviously holds for any $q = 1$ and $j = 0$. Indeed, since $\left(\phi_{k}\right)_{k \geq 0}$ is a complete orthonormal system in \(\ltwo(\mathbb{R}^m, P_{\xi})\), it holds in \(\ltwo(\mathbb{R}^m, P_{\xi_{1}})\) that
\[
f(X^x_{1})=\PE[f(X^x_{1})]+\sum_{k\geq1}a_{1,1,k}(x)\phi_{k}(\xi_{1})
\]
for any bounded $f$ with $a_{1,1,k}(x)=\PE[f(X_{1}^x)\phi_{k}(\xi_{1})]$.
Assume now that (\ref{eq:mart_repr}) holds for any $q \leq v$, $j < q$ and bounded measurable functions $f$. Let us prove that the induction assumption holds for $q=v+1$ and any $j < v+1$.
Denote for $n,k \in \nset$ and $y \in \rset^d$,
\begin{align}
a_{v,v+1,k} &= \int f \circ \Phi_v(y,e_{v+1}) \phi_k(e_{v+1}) P_\xi(\rmd e_{v+1}) \eqsp,\\
F_{n,v,1}^y &= \int f \circ \Phi_v(y,e_{v+1}) P_\xi(\rmd e_{v+1}) + \sum_{k=1}^n a_{v,v+1,k}(y) \phi_k(e_{v+1}) \eqsp.
\end{align}
The orthonormality and completeness
of the system $\left(\phi_{k}\right)_{k=0}^\infty$ implies that
\begin{equation}
\label{eq:convergence-in-ltwo}
\lim_{n \to \infty} \int | f \circ \Phi_v(y,e_{v+1}) - F_{n,v,1}^y|^2 P_\xi(\rmd e_{v+1}) \eqsp.
\end{equation}
The Parseval inequality implies that
\begin{equation}
\label{eq:convergence-bound}
\int | f \circ \Phi_v(y,e_{v+1}) - F_{n,v,1}^y|^2 P_\xi( \rmd e_{v+1})
\leq \int | f \circ \Phi_v(y,e_{v+1}) - F_{0,v,1}^y|^2 P_\xi(\rmd e_{v+1}) \leq \| f \|_{\infty}^2 \eqsp.
\end{equation}
By construction, $X_{v+1}^x= X_{v,v+1}^{X_v^x}$ and
$\int f(\Phi_v(X_v^x,e_{v+1})) P_\xi(\rmd e_{v+1})= \CPE{f(X_{v+1}^x)}{\mcg_v}$ $\P$-a.s.
Hence, using \eqref{eq:convergence-in-ltwo} and \eqref{eq:convergence-bound}, we get that
\[
\lim_{n \to \infty} \PE\Bigl[ f(X_{v+1}^x) - \CPE{f(X_{v+1}^x)}{\mcg_v} - \sum_{k=1}^n a_{v,v+1,k}(X_v^x) \phi_k(\xi_{v+1}) \Bigl| \Bigr|^2 \Bigr] = 0
\]
or equivalently
\begin{equation}
\label{eq:1-step-decomp}
f(X^x_{v+1}) = \CPE{f(X^x_{v+1})}{\mcg_v} + \sum\limits_{k=1}^{\infty}a_{v+1,v+1,k}(X^x_{v})\phi_{k}(\xi_{v+1})
\end{equation}
in $\ltwo (\rset^{mq},P^{\otimes q}_{\xi})$ which is the required statement in the case $q = v+1$ and $j=v$.
Consider now the case $q=v+1$ and $j<v$.
Set $g(y) = \int f \circ \Phi_v(y, e_{v+1}) P_\xi(\rmd e_{v+1})$. Note that $\mathsf{P}$-a.s. it holds $g(X_v^x) = \CPE{f(X_v^x)}{\mcg_v}$ and $g$ is bounded by construction. Hence, we may apply the induction hypothesis to function the bounded measurable function, which implies
\begin{equation}\label{eq:28082017a1}
\CPE{f(X^x_{v+1})}{\mcg_v}=\CPE{f(X^x_{v+1})}{\mcg_j} +\sum_{k=1}^\infty \sum_{l=j+1}^{v}a_{v+1,l,k}(X^x_{l-1})\phi_{k}(\xi_{l})
\end{equation}
with $a_{v+1,l,k}(y) = \PE[ g(X_{l-1,v}^y) \phi_k(\xi_l)]$. Using that $g(X_{l-1,v}^y)= \int f \circ \Phi_v(X_{l-1,v}^y,e_{v+1}) P_\xi(\rmd e_{v+1})$,
and $\Phi_v(G_{l-1,v}(y,e_l,\dots,e_{v}),e_{v+1})= G_{l-1,v+1}(y,e_l,\dots,e_{v+1})$, w
\begin{align*}
a_{v+1,l,k}(y) &= \PE[ g(X_{l-1,v}^y) \phi_k(\xi_l)] =
\PE[ g \circ G_{l-1,v}(y,\xi_l,\dots,\xi_v) \phi_k(\xi_l)] \\
&= \PE[ f \circ G_{l-1,v+1}(y,\xi_l,\dots,\xi_{v+1}) \phi_k(\xi_l)]= \PE[f(X^y_{l-1,v+1})\phi_{k}(\xi_{l})]
\end{align*}
Eqs.\eqref{eq:1-step-decomp}
and~\eqref{eq:28082017a1} conclude the induction step for $q = v+1$ and all $j < v+1$ and hence the proof.

\subsection{Proof of \Cref{lem:a_repr}}
\label{proof:lem:a_repr}
%For $\kind \in \nset_0^d$
%we use  a shorthand notation
%\begin{equation}
%\label{eq:definition-differential-f-p}
%\partial_{1}^{\kind} f\left(X_{p}^x\right)
%:=\partial_{z_1}^{\kind} [f\circ G_{p}](x,\chunk{Z}{1}{p}),
%\end{equation}
%provided that $f \circ G_{p}$ is smooth.
%*** Already defined
%Recall that we denote  $\boldsymbol{\varphi}(z)=\frac{1}{(2\pi)^{d/2}} \exp\{-|z|^2/2\}$,
%$z\in\mathbb R^d$ the density of the standard Gaussian $d$-dimensional random vector. For the normalized Hermite polynomial $\mathbf H_{\mathbf k}$ on $\mathbb R^d$,
%$\kind \in \mathbb N_0^d$, it holds
%$$
%\mathbf{H}_{\kind}(z)\boldsymbol{\varphi}(z)
%=\frac{(-1)^{|\kind|}}{\sqrt{\kind !}} \partial^{\kind} \boldsymbol{\varphi}(z).
%$$
Applying the integration by parts in vector form
(below $\prod_{j=l+1}^p:=1$ whenever $l \geq p$),
\begin{align*}
&\bar a_{p,\kind}(x)
 =
\idotsint_{\mathbb R^d}
[f\circ G_{p}](x,\chunk{z}{1}{p})
\mathbf{H}_{\kind}(z_{1})\boldsymbol{\varphi}(z_1)
\prod_{j=2}^p\boldsymbol{\varphi}(z_j)\, \rmd z_{1}\ldots \rmd z_{p}
\\
& =
\frac{1}{\sqrt{\kind!}}
\idotsint_{\mathbb R^d}
[f\circ G_{p}](x,\chunk{z}{1}{p})
(-1)^{|\kind|}\partial^{\kind} \boldsymbol{\varphi}(z_1)
\prod_{j=2}^p\boldsymbol{\varphi}(z_j)\, \rmd z_{1}\ldots  \rmd z_{p}
\\
& =
\frac{1}{\sqrt{\kind!}}
\idotsint
\partial_{z_1}^{\kind'}[f\circ G_{p}](x,\chunk{z}{1}{p})
(-1)^{|\kind-\kind'|}\partial^{\kind-\kind'} \boldsymbol{\varphi}(z_1)
\prod_{j=2}^p\boldsymbol{\varphi}(z_j)\, \rmd z_{1}\ldots \rmd z_{p}
\\
 & =\frac{\sqrt{(\kind-\kind')!}}{\sqrt{\kind!}}\PE\left[\partial_{z_1}^{\kind'}[f\circ G_{p}](x,\chunk{Z}{1}{p})\mathbf{H}_{\kind-\kind'}(Z_{1})\right].
\end{align*}
The last expression yields the result.

\subsection{Proof of \Cref{prop:main}}
\label{proof:prop:main}
For multi-indices $\kind,\kind^{\prime} \in \mathbb{N}_0^d$
with $\kind' \leq \kind$ componentwise
and $\kind' \ne \kind$, $\| k' \| \leq K$,
we obtain from Lemma~\ref{lem:a_repr}, that for $q \in \nset$,
\begin{equation*}
A_{q,\kind}(x)
=\left(\frac{(\kind-\kind')!}{\kind!}\right)^{1/2}
\,
\PE\left[ \sum\limits_{r=1}^{q}\{\partial_{z_1}^{\kind'} f(X^x_r)-\PE[\partial_{z_1}^{\kind'}f(X^x_r)]\}\mathbf{H}_{\kind-\kind'}(Z_1)\right],
\end{equation*}
where $A_{q,\kind}$ is defined in \eqref{eq:definition-A-s,k}. Given $\kind \in \mathbb N_0^d$,
by taking $\kind' = \kind'(\kind)
=K(\indiacc{k_1>K}\,\ldots, \indiacc{k_d>K})$, we get
\begin{multline}
\label{eq:sum_abar}
\sum_{\kind\colon\|\kind\|\geq K+1}A^2_{q,\kind}(x)
=\sum_{\kind\colon\|\kind\|\geq K+1}\left(\frac{(\kind-\kind')!}{\kind!}\right)R_q(\kind',\kind-\kind')\\
= \left\{ \sum_{I\subseteq\{1,\ldots,d\},\, I\neq \emptyset}\sum_{\mathbf{m}_{I}\in\mathbb{N}_{I}^{d}}\frac{\mathbf{m}_{I}!}{\left(\mathbf{m}_{I}+\mathbf{K}_{I}\right)!} \right\}
\left\{ \sum_{\mathbf{m}_{I^c}\in \mathbb{N}^d_{0,I^c},\,\|\mathbf{m}_{I^c}\|\leq K}R^{x}_q(\mathbf{\mathbf{K}}_{I},\mathbf{m}_{I}+\mathbf{m}_{I^c}) \right\},
\end{multline}
where for any two multi-indices \(\rind,\) \(\qind\) from \(\nset_0^d\) we have defined
\begin{equation*}
R^{x}_q(\rind,\qind)
=
\biggl\{\PE\Bigl[\sum_{p=1}^{q}\left\{\partial_{z_1}^{\rind}f\left(X^x_{p}\right)-\PE\left[\partial_{z_1}^{\rind}f\left(X^x_{p}\right)\right]\right\}\mathbf{H}_{\qind}(Z_{1})\Bigr]\biggr\}^{2}.
\end{equation*}
In \eqref{eq:sum_abar} the first sum runs over all nonempty subsets $I$ of the set $\{1,\ldots,d\}.$
For any subset $I,$ $\mathbb{N}_{I}^{d}$ stands for a set
of multi-indices $\mathbf{m}_{I}$ with elements $m_{i}=0,$ $i\not\in I,$
and $m_{i}\in\mathbb{N},$  $i\in I.$ Moreover, \(I^c=\{1,\ldots,d\}\setminus I\) and \(\mathbb{N}^d_{0,I^c}\) stands for a set
of multi-indices $\mathbf{m}_{I^c}$ with elements $m_{i}=0,$ $i\in I,$
and $m_{i}\in\mathbb{N}_0,$  $i\not\in I$. Applying the estimate
\begin{eqnarray*}
\frac{\mathbf{m}_{I}!}{\left(\mathbf{m}_{I}+\mathbf{K}_{I}\right)!}\leq (1/2)^{|I| K},
\end{eqnarray*}
we get
\begin{align}
\label{eq:sumA}
\sum_{\kind\colon\|\kind\|\geq K+1}A^2_{q,\kind}(x)
&\leq
\sum_{I\subseteq\{1,\ldots,d\},\, I\neq \emptyset} (1/2)^{|I|K} \sum_{\mathbf{m}_{I}\in\mathbb{N}_{I}^{d}} \sum_{\mathbf{m}_{I^c}\in \mathbb{N}^d_{0,I^c},\,\|\mathbf{m}_{I^c}\|\leq K} R^{x}_{q}(\mathbf{\mathbf{K}}_{I},\mathbf{m}_{I}+\mathbf{m}_{I^c})
\\
\nonumber
&\leq
\sum_{I\subseteq\{1,\ldots,d\},\, I\neq \emptyset} (1/2)^{|I|K} \sum_{\mathbf{m}\in\mathbb{N}_0^{d}} R^{x}_{q}(\mathbf{\mathbf{K}}_{I},\mathbf{m}).
\end{align}
The Parseval identity implies that for any function $\varphi: \rset^d \to \rset$ satisfying $\PE[\varphi^2(Z_1)] < \infty$,
\[
\sum_{\mathbf{m}\in \nset^d_0} \{\PE[\varphi(Z_1) \mathbf{H}_\mathbf{m}(Z_1)] \}^2 \le \PE[\{\varphi(Z_1)\}^2]
\]
Using this identity in \eqref{eq:sumA} implies
\begin{equation}
\label{eq:sumA_var_bound}
\sum_{\kind\colon\|\kind\|\geq K+1}A^2_{q,\kind}(x)
\leq \sum_{I\subseteq\{1,\ldots,d\},\, I\neq \emptyset}
\left(1/2\right)^{|I|K}
\PVar\biggl(\sum_{p=1}^{q}\partial_{z_1}^{\mathbf{K}_I}f\left(X^x_{p}\right)
\biggr).
\end{equation}
The sum
$\sum_{p=1}^{q} \partial_{z_1}^{\bigK_I} f(X_p^x)$
is  a function of $x,\chunk{Z}{1}{q}$: $\sum_{p=1}^q  \partial_{z_1}^{\mathbf{K}_I} f(X_p^x)
=F(x,\chunk{Z}{1}{q})$.
By the Gaussian Poincar\'e inequality \cite{blm:2013}, we have
\[
\mathsf{Var}\biggl[\sum\nolimits_{p=1}^{q}\partial_{z_1}^{\bigK_I} f(X_p^x)\biggr]
\le \PE \biggl[\|\nabla_z F(x,\chunk{Z}{1}{q})\|^2
\biggr],
\]
where $\nabla_z F=(\nabla_{z_1} F,\ldots,\nabla_{z_q} F)$ and $\nabla_{z_j}$ is defined in \eqref{eq:definition-gradient}. Hence,
%Notice that \(\nabla_{Z_j} F=\sqrt{\gamma}\,\nabla_{y_j} F\), hence
$$
\mathsf{Var}\biggl[\sum_{p=1}^{q}\partial_{z_1}^{\mathbf{K}_I} f(X_p^x)\biggr] \leq \sum_{j=1}^{q}\mathsf{E}\biggl[\bigl\|\nabla_{z_{j}}\sum_{p=1}^{q}\partial_{z_1}^{\mathbf{K}_I} f(X_p^x) \bigr\|^{2}\biggr].
$$
Note that $\nabla_{z_j} \partial_{z_1}^{\mathbf{K}_I} f(X_p^x) = 0$ for $p < j$. Together with \eqref{eq:sumA_var_bound} this implies the statement \eqref{eq:sumA_lemma}.

\subsection{Proof of \Cref{prop:var_bounds}}
\label{proof:prop:var_bounds}
Recall that, due to \eqref{eq:expression-variance}, for $K \in \nset$,
\begin{equation}
\label{eq:var_pi_n_K}
\PVar[\pi^{(x,K)}_{n}(f)]=\frac{1}{n^2}\sum_{\|\kind \| \geq K+1}^{\infty}\sum_{l=1}^{n}\PE[A^2_{n-l+1, \kind} (X^x_{l-1})].
\end{equation}
By \Cref{prop:main}, for fixed $q \in \nset$, and any $x \in \rset^d$,
\begin{equation}
\label{eq:var_bound}
\sum_{\|\kind \| \geq K+1}^{\infty}A^2_{q,\kind}(x)
\leq \sum\limits_{I \subseteq{1,\ldots,d}, I \neq \emptyset}(1/2)^{|I|K}
\sum_{j=1}^{q}\PE\biggl[ \bigl\| \sum_{p=j}^{q} \nabla_{z_{j}} \partial_{z_1}^{\bigK_{I}}f(X_p^x)\bigr\|^2\biggr].
\end{equation}
Now we fix $p$ and $j$ in $\{1,\ldots,q\}$, such that $p \geq j$,  and a non-empty subset $I \subseteq \{1,\ldots,d\} $. By the multivariate Fa\`a di Bruno's formula \cite{constantine:faa_di_bruno:98},
\[
\partial_{z_1}^{\bigK_I}f(X_p^x) = \sum\limits_{1 \leq |\rind| \leq K|I|}f^{(\rind)}(X_p^x)\sum\limits_{P(\bigK_I,\rind)}\frac{\bigK_I!}{\kind_1! \dots \kind_{K|I|}!}  \prod\limits_{i=1}^{K|I|}\frac{\bigl[\partial_{z_1}^{\lind_i}X_p^x\bigr]^{\kind_i}}{[\lind_i!]^{|\kind_i|}}\eqsp,
\]
where the set $P(\bigK_I,\rind)$ is defined in \Cref{par:notations}. Hence, we obtain
\begin{eqnarray*}
\nabla_{z_j} \partial_{z_1}^{\bigK_I}f(X_p^x) &=& \sum\limits_{1 \leq |\rind| \leq K|I|} \Jac_{X_p^x}^{(z_j)} \bigl\{ \nabla f^{(\rind)}(X_p^x)\bigr\}\sum\limits_{P(\bigK_I,\rind)}\frac{\bigK_I!}{\kind_1! \dots \kind_{K|I|}!}  \prod\limits_{i=1}^{K|I|}\frac{\bigl[\partial_{z_1}^{\lind_i}X_p^x\bigr]^{\kind_i}}{[\lind_i!]^{|\kind_i|}} \\
&&+ \sum\limits_{1 \leq |\rind| \leq K|I|}f^{(\rind)}(X_p^x)\sum\limits_{P(\bigK_I,\rind)}\frac{\bigK_I!}{\kind_1! \dots \kind_{K|I|}!}\sum\limits_{s=1}^{K|I|}\prod\limits_{i=1, i \neq s}^{K|I|}\frac{\bigl[\partial_{z_1}^{\lind_i}X_p^x\bigr]^{\kind_i}}{\bigl[\lind_i!\bigr]^{|\kind_i|}} \\
&& \times \frac{1}{\bigl[\lind_{s}!\bigr]^{|\kind_s|}}\sum\limits_{u=1}^{d}\prod\limits_{1 \leq v \leq d, v \neq u}^{d}\big[\partial_{z_1}^{\lind_{s}}(X_p^x)_{v}\bigr]^{\kind_{s,v}}\kind_{s,u}\big[\partial_{z_1}^{\lind_{s}}(X_p^x)_{u}\bigr]^{\kind_{s,u}-1}\nabla_{z_j}\partial_{z_1}^{\lind_s}(X_p^x)_{u}
\end{eqnarray*}
Using the bounds of \Cref{lem:main_aux_lemma_multivariate} and \Cref{lem:matrix_norm}, we obtain
\begin{eqnarray}
\label{eq:part-part}
\bigl\| \nabla_{z_{j}}\partial_{z_{1}}^{\bigK_I}f\left(X_{p}^x\right) \bigr\|
\leq \gamma^{(K|I|+1)/2} A_{K|I|} \prod_{k=2}^{p} \bigl\|\Id-\gamma J_{b}(X_{k-1}^x)\bigr\| \leq \gamma^{(K|I|+1)/2} A_{K|I|}(1-\gamma m/2)^{p-1}
\end{eqnarray}
with a suitable constant $A_{K|I|}$. Substituting into~\eqref{eq:var_bound}, we obtain
\begin{align*}
\sum_{\|\kind\| \geq K+1}^{\infty}A^2_{q,\kind}(x)
\leq
\sum\limits_{I \subseteq \{1,\dots,d\}, I \neq \emptyset}\gamma^{K|I|+1} A_{K|I|}^{2} \sum_{j=1}^{q} \left(\sum_{p=j}^{q}(1-\gamma m/2)^{p-1}\right)^{2}
\leq
\frac{\gamma^{K-2} B_{K}}{m^3(1-\gamma m/2)^2}
\end{align*}
with a constant $B_K$ not depending on $\gamma$ and $n$. Hence, due to \eqref{eq:var_pi_n_K}, we obtain
\[
\PVar[\pi^{(x,K)}_{n}(f)] \lesssim \frac{\gamma^{K-2}}{n}.
\]
For the truncated estimate $\pi_{n,\ntrunc}^{(K)}(f)$, we obtain using \eqref{eq:expression-variance-trunc} that
\begin{equation}
\label{eq:trunc_var_bound}
\PVar\bigl[\pi_{n,\ntrunc}^{(x,K)}(f)\bigr] = \PVar\bigl[\pi_{n}^{(x,K)}(f)\bigr] + \frac{1}{n^2}\sum\limits_{1 \leq \| \kind \| \leq K}\sum\limits_{l=1}^{n-\ntrunc}\mathsf{E}\biggl[\bigl(\sum\limits_{r=\ntrunc+1}^{n-l}\bar{a}_{r,\kind}(X_{l-1}^{x})\bigr)^2\biggr]
\end{equation}
Let us bound the quantity $\sum_{1 \leq \|\kind\| \leq K}\bigl[\sum_{r=\ntrunc+1}^{n-l}\bar{a}_{r,\kind}(y)\bigr]^2$ for $y \in \rset$. First, let us show that for any $r \in \nset$, any $x,y \in \rset$,
\begin{equation}
\label{eq:Q_r_bound}
\bigl| Q_r(x) - Q_r(y) \bigr|^2 \leq C_{f}^2(1- \gamma m)^{r}\|x-y\|^2\,.
\end{equation}
Indeed, it holds
\begin{eqnarray*}
\mathsf{E}\|X_{r}^x - X_{r}^y\|^2 &=& \mathsf{E}\|X_{r-1}^x - X_{r-1}^y - \gamma \bigl(b(X_{r-1}^x) - b(X_{r-1}^y)\bigr)\|^2 \\
&=&  \mathsf{E}\bigl[\|X_{r-1}^x - X_{r-1}^y\|^2 - 2\gamma \langle X_{r-1}^x - X_{r-1}^y, b(X_{r-1}^x) - b(X_{r-1}^y) \rangle \\
&+&  \gamma^2 \|b(X_{r-1}^x) - b(X_{r-1}^y)\|^2 \bigr] \\
&\leq& (1-\gamma m)\mathsf{E}\|X_{r-1}^x - X_{r-1}^y\|^2,
\end{eqnarray*}
provided that $\gamma \in (0, m/\ltwo)$. Then
\begin{align*}
\bigl|Q_r(x) - Q_r(y)\bigr|^2 &= \bigl|\mathsf{E}\bigl[f(X_r^{x})\bigr] - \mathsf{E}\bigl[f(X_r^{y})\bigr] \bigr|^2 \leq C^2_{f}\, \mathsf{E}\|X_r^{x} - X_r^{y}\|^2 \\
&\leq C_{f}^2 (1-\gamma m) \mathsf{E}\|X_{r-1}^{x} - X_{r-1}^{y}\|^2
\end{align*}
and \eqref{eq:Q_r_bound} follows. Note that, by the definition of $\bar{a}_{r,\kind}$,
\[
\sum\limits_{r=\ntrunc+1}^{n-l}\bar{a}_{r,\kind}(y) = \mathsf{E}\biggl[H_{\kind}(Z)\sum\limits_{r=\ntrunc+1}^{n-l}Q_{r-1}(\Phi(y,Z))\biggr].
\]
The Parseval identity implies that
\begin{align*}
\sum\limits_{\|\kind\| \geq 1}\biggl(\sum\limits_{r=\ntrunc+1}^{n-l}\bar{a}_{r,\kind}(y)\biggr)^2 &= \PVar\biggl[\sum\limits_{r=\ntrunc+1}^{n-l}Q_{r-1}(\Phi(y,Z))\biggr] \\
& \leq \mathsf{E}\biggl[\bigl(\sum\limits_{r=\ntrunc+1}^{n-l}Q_{r-1}(\Phi(y,Z)) - Q_{r-1}(\Phi(y,0))\bigr)^2\biggr] \\
& \leq \frac{C_f^2 (1- \gamma m)^{2\ntrunc - 2}}{m^2\gamma}
\end{align*}
Hence, selecting $\ntrunc = \ntrunc(\gamma)$ such that $(1 - \gamma m)^{2 \ntrunc - 2} \leq \gamma^{K-1}$ for $\gamma \searrow 0_{+}$, we obtain using \eqref{eq:trunc_var_bound} that
\[
\PVar\bigl[\pi_{n,\ntrunc}^{(x,K)}(f)\bigr] \lesssim \frac{\gamma^{K-2}}{n}
\]

\subsection{Auxiliary lemmas}
\begin{lemma}
\label{lem:main_aux_lemma_multivariate}
Under the assumptions of \Cref{prop:var_bounds}, for any multi-index $\kind \in \nset_0^{d}$ with $|\kind| \geq 1$, any $j \geq 1$, and $p > j$, it holds
\begin{equation}
\label{eq:derivative_bound_multidim}
\frobnorm{\nabla_{z_j} \partial_{z_1}^{\kind} X^x_p} \leq \gamma^{(|\kind|+1)/2} C_{|\kind|} \prod_{k=2}^{p} \normop{\Id - \gamma \Jac_{b}(X^{x}_{k-1})} %\quad \text{a.s.}
\end{equation}
with constant $C_{|\kind|}$ not depending on $\gamma, j,$ and $p$. For $p \leq j$, it holds
$\nabla_{z_{j}}\partial_{z_1}^{\kind}X^x_{p}=0$.
\end{lemma}

\begin{proof}
In the proof we use the notation
\begin{equation}
\label{eq:definition-alpha}
\alpha_k := \normop{\Id -\gamma \Jac_{b}(X_{k-1}^x)},\quad k\in\mathbb N.
\end{equation}
For $v \in \{1,\dots,d\}$, we also write $\eind_{v}$ for $v-$th coordinate basis vector, that is, $\eind_{v} \in \nset_{0}^{d}$ and $(\eind_{v})_{i} = \indiacc{i}(v)$ for $i \in \{1,\dots,d\}$.
\par
We preface the lemma by some elementary but useful identities. For any multi-index $\kind$ with $|\kind| = 1$, any $i < p$, it holds
\begin{equation}
\label{eq:derivative_z_i_multidim}
\partial^{\kind}_{z_{i}}X^x_{p}
=\left[\Id-\gamma \Jac_{b}(X^x_{p-1})\right]\partial^{\kind}_{z_{i}}X^x_{p-1}.
\end{equation}

\begin{equation}
\label{eq:derivative_z_i_multidim_recurrent}
\partial^{\kind}_{z_{i}}X^x_{p}
=\sqrt{\gamma} \prod\limits_{k=i+1}^{p}\left[\Id - \gamma \Jac_{b}(X^x_{k-1})\right] \eind_{\kind}, \quad \nabla_{z_{p}}X^x_{p} = \sqrt{\gamma}\,\Id
\end{equation}
Since $X_p^x = G_p(x,Z_{1:p})$, obviously $\nabla_{z_{j}}\partial_{z_1}^{\kind}X^x_{p} = 0$ for $p < j$. For $p=j$, the statement of the lemma follows from \eqref{eq:derivative_z_i_multidim_recurrent}. Now we consider the case $p > j$. Fix $j \in \nset$ and prove~\eqref{eq:derivative_bound_multidim} for all $p>j$ by induction in~$|\kind|$. We start from $|\kind| = 1$.
%Hence, for a given $j$,
Note that for a given index $j$ and $u, v \in \{1,\dots,d\}$, the relation \eqref{eq:derivative_z_i_multidim} implies
\[
\partial^{\eind_v}_{z_j}\partial^{\kind}_{z_1}(X_p^x)_{u} = \sum\limits_{l=1}^{d}\bigl(\delta_{u,l} - \gamma b^{(\eind_l)}_{u}(X_{p-1}^{x})\bigr)\partial_{z_j}^{\eind_v}\partial_{z_1}^{\kind}(X_{p-1}^{x})_{l} - \gamma \ps{D^2 b_u(X_{p-1}^{x})\partial^{\eind_{v}}_{z_{j}}X_{p-1}^{x}}{\partial^{\kind}_{z_1}X_{p-1}^{x}}
%\gamma \sum\limits_{l=1}^{d}\sum\limits_{r=1}^{d}\frac{\partial^{2}b_{u}(X_{p-1}^{x})}{\partial_{x_r}\partial_{x_l}} \partial^{\eind_{v}}_{z_{i}}(X_{p-1}^{x})_{r}\partial^{\kind}_{z_{1}}(X_{p-1}^{x})_{l}
\]
Hence, we can write that
\begin{equation}
\label{eq:recurrence}
\nabla_{z_{j}}\partial^{\kind}_{z_{1}}X^x_{p}
=\left[\Id - \gamma \Jac_{b}(X^x_{p-1})\right]\nabla_{z_{j}}\partial^{\kind}_{z_{1}}X^x_{p-1}-
\gamma H(X^x_{p-1}), %b^{\prime\prime}(X^x_{p-1})\partial_{z_{j}}X^x_{p-1}\partial_{z_{1}}X^x_{p-1}.
\end{equation}
where the matrix $H(X^x_{p-1}) \in \rset^{d \times d}$, with the entries
\[
H(X^x_{p-1})_{u,v} =
\ps{D^2 b_u(X_{p-1}^{x})\partial^{\eind_{v}}_{z_{j}}X_{p-1}^{x}}{\partial^{\kind}_{z_1}X_{p-1}^{x}}\eqsp.
%\sum\limits_{l=1}^{d}\sum\limits_{r=1}^{d}\frac{\partial^{2}b_{u}(X_{p-1}^{x})}{\partial_{x_r}\partial_{x_l}} \partial^{\eind_{v}}_{z_{i}}(X_{p-1}^{x})_{r}\partial^{\kind}_{z_{1}}(X_{p-1}^{x})_{l}.
\]
The recurrence \eqref{eq:recurrence} implies that
\begin{equation}
\label{eq:frobnorm}
\frobnorm{\nabla_{z_{j}}\partial^{\kind}_{z_{1}}X^x_{p}} \leq \normop{\Id - \gamma \Jac_{b}(X_{p-1}^{x})} \frobnorm{\nabla_{z_{j}}\partial^{\kind}_{z_{1}}X^x_{p-1}} + \gamma \frobnorm{H(X^x_{p-1})}\eqsp.
\end{equation}

To bound $\frobnorm{H(X^x_{p-1})}$\,, we observe
\begin{align*}
\bigl|H(X^x_{p-1})_{u,v}\bigr| & \leq \frobnorm{D^2b_u(X_{p-1}^{x})}\normop{\partial^{\kind}_{z_{1}}X^x_{p-1}}
\normop{\partial^{\eind_{v}}_{z_{j}}X^x_{p-1}} \\
 & \leq \gamma d C_b \prod_{k=2}^{p-1}\alpha_k\prod_{k=j+1}^{p-1}\alpha_k
\end{align*}

Hence, using \eqref{eq:derivative_z_i_multidim_recurrent} and \eqref{eq:frobnorm}, we obtain
\begin{align*}
\frobnorm{\nabla_{z_{j}}\partial^{\kind}_{z_{1}}X^x_{p}}
&\leq \alpha_p \frobnorm{\nabla_{z_{j}}\partial^{\kind}_{z_{1}}X^x_{p-1}}
+\gamma^2 d^2 C_b\prod_{k=2}^{p-1}\alpha_k
\prod_{k=j+1}^{p-1}\alpha_k\\
&\le\alpha_p \frobnorm{\nabla_{z_{j}}\partial^{\kind}_{z_{1}}X^x_{p-1}}
+\gamma^2 d^2 C_b \prod_{k=2}^{j}\alpha_k \prod_{k=j+1}^{p-1}\alpha_k^2\eqsp.%, \quad p\ge j+1\,.
\end{align*}
%Note that $\nabla_{z_j}\partial^{\kind}_{z_1}X^x_j=0$
Now we can apply \Cref{lem:aux_lemma_products_multidim} to bound $\frobnorm{\nabla_{z_{j}}\partial^{\kind}_{z_{1}}X^x_{p}}$ with $\C_1 = \gamma^2 d^2 C_b\prod_{k=2}^{j}\alpha_k$, and, using \Cref{cor:gamma_decay}, we obtain for all $\gamma \in (0,m/\ltwo)$, that
\[
\frobnorm{\nabla_{z_{j}}\partial^{\kind}_{z_{1}}X^x_{p}} \leq \frac{2\gamma}{m}C_bd^2 \prod\limits_{k=2}^{p}\alpha_{k}\eqsp,
\]
which imply \eqref{eq:derivative_bound_multidim} for any multi-index $\kind$ with $|\kind| = 1$ with the constant $C_{1} = 2C_bd^2/m$.
\par
The induction hypothesis is therefore that the inequality
\begin{equation}
\label{eq:induc-derivative_multivar}
\frobnorm{\nabla_{z_j} \partial_{z_1}^{\qind} X^x_p} \leq \gamma^{(|\qind|+1)/2} C_{|\qind|}  \prod_{k=2}^{p}\alpha_k %\normop{\Id - \gamma \Jac_{b}(X^{x}_{k-1})}
\end{equation}
holds for all multi-indices $\qind$ with $|\qind| < r \leq Kd$ and $p>j$. We need to show~\eqref{eq:induc-derivative_multivar} for all multi-indices $\qind$ with $|\qind| = r$.
The multivariate Fa\`a di Bruno's formula \cite{constantine:faa_di_bruno:98} implies for $|\qind| \geq 2$, $p>1$ and $u \in \{1,\dots,d\}$, that
\begin{equation}
\label{eq:faa_di_bruno_multidim}
\partial_{z_1}^{\qind}(X_p^x)_{u} = \sum\limits_{1 \leq |\rind| \leq |\qind|} \biggl(\bigl(X_{p-1}^{x})_{u} - \gamma b_{u}(X_{p-1}^{x})\biggr)^{(\rind)}\sum\limits_{P(\qind,\rind)}\frac{\qind!}{\kind_1! \dots \kind_{|\qind|}!}  \prod\limits_{i=1}^{|\qind|}\frac{\bigl[\partial_{z_1}^{\lind_i}X_{p-1}^x\bigr]^{\kind_i}}{[\lind_i!]^{|\kind_i|}}.
\end{equation}
Here $\bigl[\partial_{z_1}^{\lind_i}X_{p-1}^x\bigr]^{\kind_i}$ is defined in \eqref{eq:vector_multiderivative}, and the summation is taken over the set $P(\qind,\rind)$ of multi-indices $\kind_{i}, \lind_i \in \nset_{0}^{d}$,
%\[
%P(\qind,\rind) = \{(\kind_{1},\dots,\kind_{|\qind|},\lind_{1},\dots,\lind_{|\qind|})\},
%\]
such that for some $1 \leq s \leq |\qind|$, $\kind_i = 0$ and $\lind_i = 0$ for $1 \leq i \leq |\qind| - s$, $|\kind_i| > 0$ for $|\qind|-s+1 \leq i \leq |\qind|$ and $0 \prec \ell_{|\qind|-s+1} \prec \dots \prec \ell_{|\qind|}$ are such that
\begin{equation}
\label{eq:constraints}
\sum\limits_{i=1}^{|\qind|}\kind_i = \rind, \quad \sum\limits_{i=1}^{|\qind|}|\kind_i|\lind_i = \qind.
\end{equation}
From the equation \eqref{eq:faa_di_bruno_multidim}, taking the terms with $|\rind| = 1$ out and using the fact that $(X_{p-1}^{x})^{(\rind)} = 0$ for any multi-index $\rind$ with $|\rind| \geq 2$, we have
\begin{align*}
\partial_{z_1}^{\qind}(X_p^x)_{u} &= \sum\limits_{l=1}^{d}\bigl(\delta_{u,l} - \gamma b^{(\eind_{l})}_u(X_{p-1}^{x})\bigr)\partial_{z_1}^{\qind}\bigl(X_{p-1}^{x}\bigr)_{u} \\
& - \gamma \sum\limits_{2 \leq |\rind| \leq |\qind|} b^{(\rind)}_{u}(X_{p-1}^{x})\sum\limits_{P(\qind,\rind)}\frac{\qind!}{\kind_1! \dots \kind_{|\qind|}!}  \prod\limits_{i=1}^{|\qind|}\frac{\bigl[\partial_{z_1}^{\lind_i}X_{p-1}^x\bigr]^{\kind_i}}{[\lind_i!]^{|\kind_i|}}.
\end{align*}
For $p>j$ and fixed $v \in \{1,\dots,d\}$, we then have
\begin{equation}
\label{eq:derivative_multidim_recurrence}
\partial^{\eind_{v}}_{z_{j}}\partial_{z_{1}}^{\qind}(X^x_{p})_{u}
= \sum\limits_{l=1}^{d}\bigl(\delta_{u,l} - \gamma b^{(\eind_{l})}_u(X_{p-1}^{x})\bigr)\partial_{z_j}^{\eind_{v}}\partial_{z_1}^{\qind}\bigl(X_{p-1}^{x}\bigr)_{u}  + \epsilon_{j,p}
\end{equation}
with% the quantity
\begin{align*}
\epsilon_{j,p} &= - \gamma \langle D^2b_{u}(X_{p-1}^{x}) \partial_{z_j}^{\eind_v}X_{p-1}^{x},\partial_{z_1}^{\qind}X_{p-1}^{x}\rangle \notag\\
&\hspace{2em}- \gamma \sum\limits_{2 \leq |\rind| \leq |\qind|}\biggl\{\sum\limits_{l=1}^{d}b^{(\rind + \eind_{l})}_{u}(X_{p-1}^{x})\partial_{z_j}^{\eind_v}\bigl(X_{p-1}^{x}\bigr)_{l}\biggr\}\sum\limits_{P(\qind,\rind)}\frac{\qind!}{\kind_1! \dots \kind_{|\qind|}!}  \prod\limits_{i=1}^{|\qind|}\frac{\bigl[\partial_{z_1}^{\lind_i}X_{p-1}^x\bigr]^{\kind_i}}{[\lind_i!]^{|\kind_i|}} \notag\\
&\hspace{2em}- \gamma \sum\limits_{2 \leq |\rind| \leq |\qind|}b^{(\rind)}_{u}(X_{p-1}^{x})\sum\limits_{P(\qind,\rind)}\frac{\qind!}{\kind_1! \dots \kind_{|\qind|}!}\partial_{z_j}^{\eind_{v}}\biggl[  \prod\limits_{i=1}^{|\qind|}\frac{\bigl[\partial_{z_1}^{\lind_i}X_{p-1}^x\bigr]^{\kind_i}}{[\lind_i!]^{|\kind_i|}}\biggr]
%\notag\\
%&\hspace{2em} =: \sum\limits_{l=1}^{d}\bigl(\delta_{u,l} - \gamma b^{(\eind_l)}_u(X_{p-1}^{x})\bigr)\partial_{z_j}^{\eind_{v}}\partial_{z_1}^{\qind}\bigl(X_{p-1}^{x}\bigr)_{u} + \epsilon_{j,p}
\end{align*}
%where the last equality defines the quantity $\epsilon_{j,p}$.
Furthermore,
\begin{align*}
&\partial_{z_j}^{\eind_{v}}\biggl[  \prod\limits_{i=1}^{|\qind|}\frac{\bigl[\partial_{z_1}^{\lind_i}X_{p-1}^x\bigr]^{\kind_i}}{[\lind_i!]^{|\kind_i|}}\biggr] = \sum\limits_{s=1}^{|\qind|}\biggl\{\prod\limits_{i=1, i \neq s}^{|\qind|}\frac{\bigl[\partial_{z_1}^{\lind_i}X_{p-1}^x\bigr]^{\kind_i}}{\bigl[\lind_i!\bigr]^{|\kind_i|}}\biggr\} \frac{1}{\bigl[\lind_{s}!\bigr]^{|\kind_s|}} \\
&\hspace{1em} \times
\sum\limits_{l=1}^{d}\prod\limits_{m = 1, m \neq l}^{d}\big[\partial_{z_1}^{\lind_{s}}(X_{p-1}^x)_{m}\bigr]^{\kind_{s,m}}\kind_{s,l}\big[\partial_{z_1}^{\lind_{s}}(X_{p-1}^x)_{l}\bigr]^{\kind_{s,l}-1}\partial^{\eind_v}_{z_j}\partial_{z_1}^{\lind_s}(X_{p-1}^x)_{l}.
\end{align*}
Note that the condition \eqref{eq:constraints} implies that
\[
\sum\limits_{i=1}^{|\qind|}|\kind_{i}| = |\rind|, \quad \sum\limits_{i=1}^{|\qind|}|\kind_{i}||\lind_i| = |\qind|.
\]
%Using \Cref{lem:aux_lemma_products_multidim} and
With the induction hypothesis~\eqref{eq:induc-derivative_multivar}, we bound $|\epsilon_{j,p}|$ as follows
\begin{align*}
&\left|\epsilon_{j,p}\right|  \leq  \gamma^{(|\qind|+3)/2}C_{b}C_{|\qind|-1}\prod_{k=2}^{p-1}\alpha_{k}\prod_{k=j+1}^{p-1}\alpha_{k} \\
& + \gamma^{3/2} d C_b \prod\limits_{k=j+1}^{p-1}\alpha_k \sum\limits_{2 \leq |\rind| \leq |\qind|}\sum\limits_{P(\qind,\rind)}\frac{\qind!}{\kind_1! \dots \kind_{|\qind|}!}  \prod\limits_{i=1}^{|\qind|}\frac{\prod\limits_{s=1}^{d}\bigl[C_{|\lind_{i}|}\gamma^{|\lind_i|/2}\prod\limits_{k=2}^{p-1}\alpha_k\bigr]^{\kind_{i,s}}}{[\lind_i!]^{|\kind_i|}} \\
&+ \gamma C_b \sum\limits_{2 \leq |\rind| \leq |\qind|}\sum\limits_{P(\qind,\rind)}\frac{\qind!}{\kind_1! \dots \kind_{|\qind|}!}\sum\limits_{s=1}^{|\qind|}\biggl\{\prod\limits_{i=1, i \neq s}^{|\qind|}\frac{\prod\limits_{m=1}^{d}\bigl[C_{|\lind_{i}|}\gamma^{|\lind_i|/2}\prod\limits_{k=2}^{p-1}\alpha_k\bigr]^{\kind_{i,m}}}{\bigl[\lind_i!\bigr]^{|\kind_i|}}\biggr\} \frac{1}{\bigl[\lind_{s}!\bigr]^{|\kind_s|}} \\
&\times
\sum\limits_{l=1}^{d}\prod\limits_{m = 1, m \neq l}^{d}\big[C_{|\lind_{s}|}\gamma^{|\lind_s|/2}\prod_{k=2}^{p-1}\alpha_k\bigr]^{\kind_{s,m}}\kind_{s,l}\big[C_{|\lind_s|}\gamma^{|\lind_s|/2}\prod_{k=2}^{p-1}\alpha_k\bigr]^{\kind_{s,l}-1} \\
&\times C_{|\lind_s|}\gamma^{(|\lind_s|+1)/2}\prod_{k=2}^{p}\alpha_{k}
\end{align*}
Due to \cite[Corollary 2.9]{constantine:faa_di_bruno:98},
\[
\sum\limits_{|\rind| = l}\sum\limits_{P(\qind,\rind)}\frac{\qind!}{\kind_1! \dots \kind_{|\qind|}!}\prod\limits_{i=1}^{|\qind|}\frac{1}{\bigl[\lind_{i}!\bigr]^{|\kind_{i}|}} = d^{l}S_{|\qind|}^{l}
\]
where $S_{|\qind|}^{l}$ is a Stirling number of a second kind (see \citet{constantine:combinatorics:book}). Hence, we can bound
\begin{equation}
\label{eq:epsilon_bound}
\left|\epsilon_{j,p}\right| \leq \gamma^{(|\qind|+3)/2} \, \const \prod_{k=2}^{j}\alpha_{k}\prod_{k=j+1}^{p}\alpha_{k}^{2}
\end{equation}
with some constant $\const$ depending on
$d, C_b,|\qind|,C_1,\ldots,C_{|\qind|-1}$. Thus, \eqref{eq:derivative_multidim_recurrence} and \eqref{eq:epsilon_bound} imply
$$
\frobnorm{\nabla_{z_{j}}\partial_{z_{1}}^{\qind}X^x_{p}}
\le \alpha_p \frobnorm{\nabla_{z_{j}}\partial_{z_{1}}^{\qind}X^x_{p-1}}
+\gamma^{(|\qind|+3)/2}\, \const\, \prod_{k=2}^{j}\alpha_{k}\prod_{k=j+1}^{p}\alpha_{k}^{2}, \quad p\ge j+1.
$$
We can again apply \Cref{lem:aux_lemma_products_multidim} and \Cref{cor:gamma_decay} to bound $\frobnorm{\nabla_{z_{j}}\partial_{z_{1}}^{\qind} X^x_{p}}$, and obtain~\eqref{eq:induc-derivative_multivar} for all multi-indices $\qind$ with $|\qind|=r$.
This concludes the proof.
\end{proof}

\begin{lemma}
\label{lem:aux_lemma_products_multidim}
Let $(x_p)_{p\in\mathbb N_0}$ and $(\epsilon_p)_{p\in\mathbb N}$ be sequences of nonnegative real numbers with $x_0 = 0$, satisfying
\begin{equation}
\label{eq:x_p_recur}
0 \le x_p\le\alpha_p x_{p-1} + \epsilon_p, \quad
0 \le \epsilon_p \le \C_1\prod_{k=1}^p \alpha_k^2,\quad p\in\mathbb N,
\end{equation}
for any $p\in\mathbb N$, and $\C_1$ is some nonnegative constant.
%Assume
%\begin{equation}
%\label{eq:prod_alpha_constraint}
%\sum_{r=1}^{\infty}\prod_{k=1}^{r}\alpha_k < \infty\eqsp,
%\end{equation}
Then
$$
x_p \leq  \C_1 \prod_{k=1}^p \alpha_k \biggl(\sum_{r=1}^{\infty}  \prod_{k=1}^r\alpha_k\biggr),\quad p\in\mathbb N.
$$
\end{lemma}
\begin{proof}
Applying~\eqref{eq:x_p_recur} recursively, we get
$
x_p \le
%x_0 \prod_{k=1}^p \alpha_k
\sum_{r=1}^p \epsilon_r
\prod_{k=r+1}^p \alpha_k
%= \sum_{r=1}^p \epsilon_r\prod_{k=r+1}^p \alpha_k,
$
where we use the convention $\prod_{k=p+1}^p:=1$. The proof is completed by using an upper bound on $\epsilon_{p}$.
%Using an upper bound on $\epsilon_{p}$, we obtain
%$$
%x_p \leq \C_1 \prod_{k=1}^p\alpha_k \biggl( \sum_{r=1}^{p}  \prod_{k=1}^r\alpha_k \biggr),
%$$
%which %together with~\eqref{eq:prod_alpha_constraint},
%completes the proof.
\end{proof}

\begin{lemma}
\label{lem:matrix_norm}
Assume that there exist $m > 0$, such that for any $x \in \rset^d$, $x^{\top} Ax \geq m\|x\|^2$. Then for any $\gamma \in (0,m/\|A\|^2)$, it holds
\[
\normop{\Id - \gamma A} \leq 1 - \gamma m/2\eqsp.
\]
\end{lemma}
\begin{proof}
Note that for $\gamma \in (0,m/\|A\|^2)$,
\begin{align*}
\normop{\Id - \gamma A}^2 &= \sup\limits_{x \in \rset^d}\frac{\|(\Id - \gamma A)x\|^2}{\|x\|^2} = \sup\limits_{x \in \rset^d}\frac{\|x\|^2 - \gamma x^{\top}(A+A^{\top})x + \gamma^2 x^{\top} A^{\top} A x}{\|x\|^2}  \\
& \leq 1 - 2\gamma m + \gamma^2 \|A\|^2 \leq 1 - m\gamma \eqsp.
\end{align*}
\end{proof}
\begin{corollary}
\label{cor:gamma_decay}
Under the assumptions of \Cref{prop:main}, for all $\gamma \in (0,m/\ltwo)$, it holds
\[
\gamma \sum\limits_{r=1}^{\infty}\prod_{k=1}^{r}\normop{\Id - \gamma \jac{b}(X_{k-1}^{x})} \leq \frac{2}{m}
\]
\end{corollary}

\section*{Acknowledgement} 
The publication was supported by the grant for research centers in the field of AI provided by the Analytical Center for the Government of the Russian Federation (ACRF) in accordance with the agreement on the provision of subsidies (identifier of the agreement 000000D730321P5Q0002) and the agreement with HSE University  No. 70-2021-00139.

\newpage

\bibliography{refs-1}

\begin{thebibliography}{33}
\providecommand{\natexlab}[1]{#1}
\providecommand{\url}[1]{\texttt{#1}}
\expandafter\ifx\csname urlstyle\endcsname\relax
  \providecommand{\doi}[1]{doi: #1}\else
  \providecommand{\doi}{doi: \begingroup \urlstyle{rm}\Url}\fi

\bibitem[Assaraf and Caffarel(1999)]{assaraf1999zero}
R.~Assaraf and M.~Caffarel.
\newblock Zero-variance principle for {M}onte {C}arlo algorithms.
\newblock \emph{Physical review letters}, 83\penalty0 (23):\penalty0 4682,
  1999.

\bibitem[Belomestny et~al.(2018)Belomestny, H{\"a}fner, and
  Urusov]{belomestny2018stratified}
D.~Belomestny, S.~H{\"a}fner, and M.~Urusov.
\newblock Variance reduction for discretised diffusions via regression.
\newblock \emph{Journal of Mathematical Analysis and Applications},
  458:\penalty0 393--418, 2018.

\bibitem[Belomestny et~al.(2020)Belomestny, Iosipoi, Moulines, Naumov, and
  Samsonov]{belomestny2019esvm}
D.~Belomestny, L.~Iosipoi, E.~Moulines, A.~Naumov, and S.~Samsonov.
\newblock Variance reduction for markov chains with application to {MCMC}.
\newblock \emph{Statistics and Computing}, 30\penalty0 (4):\penalty0 973--997,
  2020.
\newblock \doi{10.1007/s11222-020-09931-z}.
\newblock URL \url{https://doi.org/10.1007/s11222-020-09931-z}.

\bibitem[Ben~Zineb and Gobet(2013)]{GobetCV}
T.~Ben~Zineb and E.~Gobet.
\newblock Preliminary control variates to improve empirical regression methods.
\newblock \emph{Monte Carlo Methods Appl.}, 19\penalty0 (4):\penalty0 331--354,
  2013.
\newblock ISSN 0929-9629.
\newblock \doi{10.1515/mcma-2013-0015}.
\newblock URL \url{https://doi.org/10.1515/mcma-2013-0015}.

\bibitem[Bortoli and Durmus(2020)]{debortoli2020}
V.~D. Bortoli and A.~Durmus.
\newblock Convergence of diffusions and their discretizations: from continuous
  to discrete processes and back.
\newblock 2020.

\bibitem[Boucheron et~al.(2013)Boucheron, Lugosi, and Massart]{blm:2013}
S.~Boucheron, G.~Lugosi, and P.~Massart.
\newblock \emph{Concentration inequalities: A nonasymptotic theory of
  independence}.
\newblock Oxford University Press, 2013.

\bibitem[Brosse et~al.(2018)Brosse, Durmus, Meyn, and
  Moulines]{brosse2018diffusion}
N.~Brosse, A.~Durmus, S.~Meyn, and E.~Moulines.
\newblock Diffusion approximations and control variates for {MCMC}.
\newblock \emph{arXiv preprint arXiv:1808.01665}, 2018.

\bibitem[Constantine(1987)]{constantine:combinatorics:book}
G.~M. Constantine.
\newblock \emph{Combinatorial Theory and Statistical Design}.
\newblock Wiley, New York, 1987.

\bibitem[Constantine and Savits(1996)]{constantine:faa_di_bruno:98}
G.~M. Constantine and T.~H. Savits.
\newblock A multivariate faa di bruno formula with applications.
\newblock \emph{Transactions of the American Mathematical Society},
  348\penalty0 (2):\penalty0 503--520, 1996.
\newblock ISSN 00029947.
\newblock URL \url{http://www.jstor.org/stable/2155187}.

\bibitem[Dalalyan(2017)]{dalalyan2017theoretical}
A.~S. Dalalyan.
\newblock Theoretical guarantees for approximate sampling from smooth and
  log-concave densities.
\newblock \emph{Journal of the Royal Statistical Society: Series B (Statistical
  Methodology)}, 79\penalty0 (3):\penalty0 651--676, 2017.

\bibitem[Dellaportas and Kontoyiannis(2012)]{dellaportas2012control}
P.~Dellaportas and I.~Kontoyiannis.
\newblock Control variates for estimation based on reversible {M}arkov chain
  monte carlo samplers.
\newblock \emph{Journal of the Royal Statistical Society: Series B (Statistical
  Methodology)}, 74\penalty0 (1):\penalty0 133--161, 2012.

\bibitem[Dimov(2008)]{dimov2008monte}
I.~T. Dimov.
\newblock \emph{Monte Carlo methods for applied scientists}.
\newblock World Scientific, 2008.

\bibitem[Douc et~al.(2018)Douc, Moulines, Priouret, and Soulier]{moulines2018}
R.~Douc, E.~Moulines, P.~Priouret, and P.~Soulier.
\newblock \emph{{M}arkov Chains}.
\newblock Springer New York, 2018.

\bibitem[Durmus and Moulines(2017)]{durmus:moulines:2017}
A.~Durmus and E.~Moulines.
\newblock Nonasymptotic convergence analysis for the unadjusted {L}angevin
  algorithm.
\newblock \emph{Ann. Appl. Probab.}, 27\penalty0 (3):\penalty0 1551--1587,
  2017.
\newblock ISSN 1050-5164.
\newblock \doi{10.1214/16-AAP1238}.
\newblock URL \url{https://doi.org/10.1214/16-AAP1238}.

\bibitem[Glasserman(2013)]{glasserman2013monte}
P.~Glasserman.
\newblock \emph{Monte Carlo methods in financial engineering}, volume~53.
\newblock Springer Science \& Business Media, 2013.

\bibitem[Gobet(2016)]{GobetBook}
E.~Gobet.
\newblock \emph{Monte-{C}arlo methods and stochastic processes}.
\newblock CRC Press, Boca Raton, FL, 2016.
\newblock ISBN 978-1-4987-4622-9.
\newblock From linear to non-linear.

\bibitem[Gy{\"o}rfi et~al.(2006)Gy{\"o}rfi, Kohler, Krzyzak, and
  Walk]{gyorfi2006distribution}
L.~Gy{\"o}rfi, M.~Kohler, A.~Krzyzak, and H.~Walk.
\newblock \emph{A distribution-free theory of nonparametric regression}.
\newblock Springer Science \& Business Media, 2006.

\bibitem[Heinrich and Sindambiwe(1999)]{heinrich1999monte}
S.~Heinrich and E.~Sindambiwe.
\newblock Monte carlo complexity of parametric integration.
\newblock \emph{Journal of Complexity}, 15\penalty0 (3):\penalty0 317--341,
  1999.

\bibitem[Henderson(1997)]{henderson1997variance}
S.~G. Henderson.
\newblock \emph{Variance reduction via an approximating {M}arkov process}.
\newblock PhD thesis, Stanford University, 1997.

\bibitem[Henderson and Simon(2004)]{henderson2004}
S.~G. Henderson and B.~Simon.
\newblock Adaptive simulation using perfect control variates.
\newblock \emph{J. Appl. Probab.}, 41\penalty0 (3):\penalty0 859--876, 09 2004.
\newblock \doi{10.1239/jap/1091543430}.
\newblock URL \url{http://dx.doi.org/10.1239/jap/1091543430}.

\bibitem[Lamberton and Pagès(2002)]{lamberton:pages:2002}
D.~Lamberton and G.~Pagès.
\newblock {Recursive computation of the invariant distribution of a diffusion}.
\newblock \emph{Bernoulli}, 8\penalty0 (3):\penalty0 367 -- 405, 2002.
\newblock \doi{bj/1078779875}.
\newblock URL \url{https://doi.org/}.

\bibitem[Lemaire(2007)]{MR2353037}
V.~Lemaire.
\newblock An adaptive scheme for the approximation of dissipative systems.
\newblock \emph{Stochastic Process. Appl.}, 117\penalty0 (10):\penalty0
  1491--1518, 2007.
\newblock ISSN 0304-4149.
\newblock \doi{10.1016/j.spa.2007.02.004}.
\newblock URL \url{https://doi.org/10.1016/j.spa.2007.02.004}.

\bibitem[Liu and Zhao(2019)]{lotka_volterra_control}
J.~Liu and W.~Zhao.
\newblock Dynamic analysis of stochastic {L}otka–{V}olterra predator-prey
  model with discrete delays and feedback control.
\newblock \emph{Complexity}, 2019:\penalty0 1--15, 11 2019.
\newblock \doi{10.1155/2019/4873290}.

\bibitem[Mao et~al.(2003)Mao, Sabanis, and Renshaw]{sabanis_lv}
X.~Mao, S.~Sabanis, and E.~Renshaw.
\newblock Asymptotic behaviour of the stochastic {L}otka–{V}olterra model.
\newblock \emph{Journal of Mathematical Analysis and Applications},
  287\penalty0 (1):\penalty0 141 -- 156, 2003.
\newblock ISSN 0022-247X.
\newblock \doi{https://doi.org/10.1016/S0022-247X(03)00539-0}.
\newblock URL
  \url{http://www.sciencedirect.com/science/article/pii/S0022247X03005390}.

\bibitem[Mattingly et~al.(2002)Mattingly, Stuart, and
  Higham]{mattingly:stuart:higham:2002}
J.~Mattingly, A.~Stuart, and D.~Higham.
\newblock Ergodicity for sdes and approximations: locally lipschitz vector
  fields and degenerate noise.
\newblock \emph{Stochastic Processes and their Applications}, 101\penalty0
  (2):\penalty0 185--232, 2002.
\newblock ISSN 0304-4149.
\newblock \doi{https://doi.org/10.1016/S0304-4149(02)00150-3}.
\newblock URL
  \url{https://www.sciencedirect.com/science/article/pii/S0304414902001503}.

\bibitem[Mengersen and Tweedie(1996)]{mengersen:tweedie:1996}
K.~Mengersen and R.~L. Tweedie.
\newblock Rates of convergence of the {H}astings and {M}etropolis algorithms.
\newblock \emph{Ann. Statist.}, 24:\penalty0 101--121, 1996.

\bibitem[Mira et~al.(2013)Mira, Solgi, and Imparato]{mira2013zero}
A.~Mira, R.~Solgi, and D.~Imparato.
\newblock Zero variance markov chain {M}onte {C}arlo for bayesian estimators.
\newblock \emph{Statistics and Computing}, 23\penalty0 (5):\penalty0 653--662,
  2013.

\bibitem[Oates et~al.(2016)Oates, Girolami, and
  Chopin]{oates:girolami:chopin:2016}
C.~J. Oates, M.~Girolami, and N.~Chopin.
\newblock Control functionals for {M}onte {C}arlo integration.
\newblock \emph{Journal of the Royal Statistical Society: Series B (Statistical
  Methodology)}, pages n/a--n/a, 2016.
\newblock ISSN 1467-9868.
\newblock \doi{10.1111/rssb.12185}.
\newblock URL \url{http://dx.doi.org/10.1111/rssb.12185}.

\bibitem[Oates et~al.(2017)Oates, Girolami, and Chopin]{oates2017control}
C.~J. Oates, M.~Girolami, and N.~Chopin.
\newblock Control functionals for monte carlo integration.
\newblock \emph{Journal of the Royal Statistical Society: Series B (Statistical
  Methodology)}, 79\penalty0 (3):\penalty0 695--718, 2017.

\bibitem[Pag\`es and Panloup(2018)]{MR3861816}
G.~Pag\`es and F.~Panloup.
\newblock Weighted multilevel {L}angevin simulation of invariant measures.
\newblock \emph{Ann. Appl. Probab.}, 28\penalty0 (6):\penalty0 3358--3417,
  2018.
\newblock ISSN 1050-5164.
\newblock \doi{10.1214/17-AAP1364}.
\newblock URL \url{https://doi.org/10.1214/17-AAP1364}.

\bibitem[Rubinstein and Kroese(2016)]{rubinstein2016simulation}
R.~Y. Rubinstein and D.~P. Kroese.
\newblock \emph{Simulation and the {M}onte {C}arlo method}, volume~10.
\newblock John Wiley \& Sons, 2016.

\bibitem[South et~al.(2018)South, Oates, Mira, and
  Drovandi]{south:mira:drovandi:2018}
L.~F. South, C.~J. Oates, A.~Mira, and C.~Drovandi.
\newblock Regularised zero-variance control variates.
\newblock \emph{arXiv preprint arXiv:1811.05073}, 2018.

\bibitem[South et~al.(2021)South, Riabiz, Teymur, Oates, et~al.]{south2021post}
L.~F. South, M.~Riabiz, O.~Teymur, C.~Oates, et~al.
\newblock Post-processing of mcmc.
\newblock \emph{arXiv preprint arXiv:2103.16048}, 2021.

\end{thebibliography}

\end{document}